\documentclass[a4paper,11pt]{article}
\pdfoutput=1 % if your are submitting a pdflatex (i.e. if you have
\usepackage{jheppub} % for details on the use of the package, please
\usepackage[T1]{fontenc} % if needed

\usepackage{amssymb}
\usepackage{ulem}
\usepackage{physics}
\usepackage[utf8]{inputenc} %utf8
\usepackage{graphicx}
\usepackage{braket}
\usepackage[vcentermath]{youngtab}
\usepackage[table,xcdraw]{xcolor}
 \usepackage{longtable}
 \usepackage{fancyhdr}
\usepackage{booktabs} 
\usepackage{mathrsfs}
\usepackage[T1]{fontenc}
\usepackage{setspace}
\usepackage{amsfonts}
\usepackage{amssymb}
\usepackage{amsmath}
\usepackage{epsfig}
\usepackage{latexsym}
\usepackage{color}
\usepackage{nicematrix}
\usepackage{nicefrac}
 \usepackage{slashed}
 \usepackage{multirow}
 \usepackage{comment}
 \usepackage{soul}
 \usepackage{hyperref}
\usepackage{slashed}
\usepackage{simpler-wick}
\usepackage{array}
\usepackage{tabu}
\usepackage{tcolorbox}

\usepackage{ytableau}
\usepackage{youngtab}
\usepackage{listings}
\usepackage[english]{babel}
\usepackage[utf8]{inputenc}
\usepackage[T1]{fontenc}
\usepackage{bold-extra}
\usepackage{subfig}
\usepackage[subfigure]{tocloft}
\usepackage{comment}
\usepackage{amsfonts, amsmath, amsthm, amssymb}
\usepackage{amscd}
\usepackage{amsfonts}
\usepackage{anonchap}
\usepackage{mathtools}
\usepackage[hang, flushmargin]{footmisc}
\usepackage{footnotebackref}
 \usepackage{float}
\usepackage{physics}
\usepackage{pdfpages}
\usepackage{placeins}
\usepackage{slashed}
\usepackage[title,toc,titletoc]{appendix}
\usepackage{titlesec}
\usepackage{tocloft}
\usepackage{bbm}
\usepackage{bbold}
\usepackage{tikz-cd}
\definecolor{rossoCP3}{cmyk}{0,.88,.77,.40}
\usetikzlibrary{decorations.pathmorphing}

\newcommand{\ea}[1]{
\begin{align}
#1
\end{align}
}

\newcommand{\ee}{\end{equation}}
\newcommand{\be}{\begin{equation}}
\newcommand{\bea}{\begin{align}}
\newcommand{\eea}{\end{alig}}

\newcommand   \cO {\mathcal{O}}

\newcommand{\identity}{\mathbbm{1}}

  \definecolor{darkrosso}{RGB}{100,13,20}
\definecolor{lightor}{RGB}{248,150,30}

\def\lsim{\mathrel{\rlap{\lower4pt\hbox{\hskip1pt$\sim$}}
    \raise1pt\hbox{$<$}}}                % less than or approx. symbol
\def\gsim{\mathrel{\rlap{\lower4pt\hbox{\hskip1pt$\sim$}}
    \raise1pt\hbox{$>$}}}                % greater than or approx. symbol
    
    \def\be{\begin{equation}}
\def\ee{\end{equation}}
\def\ba{\begin{eqnarray}}
\def\ea{\end{eqnarray}}

\def\IR{\relax{\rm I\kern-.18em R}}

\def\IR{\relax{\rm I\kern-.18em R}}
\def\IL{\relax{\rm I\kern-.18em L}}

\def\inv{^{\raise.15ex\hbox{${\scriptscriptstyle -}$}\kern-.05em 1}}

\def\cO{{\cal O}}

\def\Tr{{\rm Tr}}

\title{{\textbf{On the $\theta$-angle physics of QCD under pressure:\\  The strange and isospin phase diagram}}}

\author[a]{Jahmall Bersini,}
\author[b, c, e]{Alessandra D'Alise,}
\author[c, d, e]{Clelia Gambardella, }
\author[b, c, d, e]{Francesco Sannino}

\affiliation[a]{ Kavli IPMU (WPI), UTIAS, The University of Tokyo, Kashiwa, Chiba 277-8583, Japan}
\affiliation[b]{Dipartimento di Fisica ``E. Pancini", Università di Napoli Federico II, via Cintia, 80126 Napoli, Italy}
\affiliation[c]{INFN sezione di Napoli, Complesso Universitario di Monte S. Angelo Edificio 6, via Cintia, 80126 Napoli, Italy}
\affiliation[d]{Scuola Superiore Meridionale, Largo S. Marcellino, 10, 80138 Napoli NA, Italy}
\affiliation[e]{Quantum  Theory Center ($\hbar$QTC) at IMADA and D-IAS, Southern Denmark Univ., Campusvej 55, 5230 Odense M, Denmark}

\emailAdd{jahmall.bersini@ipmu.jp}
\emailAdd{alessandra.dalise@unina.it}
\emailAdd{c.gambardella@ssmeridionale.it}
\emailAdd{sannino@qtc.sdu.dk}

\abstract{We unveil the impact of the $\theta$-angle  on the QCD phase diagram at nonzero isospin and strangeness chemical potentials for three light flavors and different quark mass ratios.  
We establish the phase boundaries as well as the nature of the associated phase transitions. The order parameters and the physical spectrum in the different phases are determined. We further elucidate the physics around $\theta=\pi$ where we discover a novel parity-preserving superfluid phase. Finally, we comment on Dashen's phenomenon in the superfluid phases when varying the number of light flavors.}

\date{}
\begin{document}

\maketitle

\section{Introduction}
\label{intro}

Quantum Chromodynamics (QCD) is a pillar of our current understanding of Nature. Nevertheless, despite the large body of work, both analytical and numerical, much is still left to be understood about its dynamics, especially in the strongly coupled regime.  
The topological sector of the theory controlled by the $\theta$-angle   \cite{Callan:1976je} plays a special role due to its theoretical and phenomenological implications. Another active and physically relevant area of research is related to the QCD dynamics under extreme conditions occurring, for example, in the core of neutron stars where matter is highly squeezed \cite{Rajagopal:2000wf}. It is therefore interesting to unveil the QCD phase diagram under the influence of the topological sector. Moreover, composite dark and bright extensions of the Standard Model often require a deeper understanding of the QCD-like strongly coupled dynamics at both nonzero $\theta$ and matter density, as it is the case of composite asymmetric dark matter as reviewed in \cite{Sannino:2009za}.  While previous studies focused primarily on the effect of $\theta$ on the $T-\mu_B$ (temperature - baryon chemical potential) phase diagram \cite{Sakai:2011gs, DElia:2013uaf, Chatterjee:2011yz, Boer:2008ct, Chatterjee:2014csa, Murgana:2024djt}, we concentrate here on the infrared dynamics of QCD at nonzero $\theta$-angle and quark chemical potentials. We mostly discuss three light quark flavors at finite isospin $\mu_I$ and strangeness $\mu_s$ chemical potentials. The study of the corresponding phase diagram at $\theta=0$ has a long history \cite{Son:2000xc, Kogut:2001id, Son:2000by, Brandt:2017zck, GomezNicola:2022asf, Ayala:2023cnt, Adhikari:2019zaj, Carignano:2016lxe, Mammarella:2015pxa, Schafer:2001bq, Abbott:2024vhj, Detmold:2008yn, Andersen:2006ys, He:2005nk, Barducci:2004tt, Loewe:2002tw, Kogut:2002zg, Adhikari:2019mlf, Mannarelli:2019hgn} and revealed the existence of two different superfluid phases that we here name {\it Pion} and {\it Kaon} phases which are respectively characterized by the emergence of pion and kaon condensation. On the other hand, the impact of the $\theta$-angle on this sector of the theory has been limited to the two flavor case\footnote{See also the related works \cite{Metlitski:2005db, Bersini:2022jhs} investigating the $\theta-\mu_B$ phase diagram in two-color QCD.} \cite{Metlitski:2005di}. While our motivation is primarily theoretical, our findings may have implications for axion physics in dense matter \cite{Balkin:2020dsr} (in particular for the relation between meson and axion condensation), the cosmological Peccei-Quinn phase transition \cite{Peccei:1977hh, Peccei:1977ur}, and strongly interacting BSM models \cite{Cacciapaglia:2020kgq}. Moreover, it has been proposed that heavy-ion collisions may generate CP-violating states characterized by an effective $\theta$ parameter \cite{Kharzeev:1998kz, Voloshin:2004vk} as well as a large isospin asymmetry eventually leading to pion condensation \cite{Ruck:1976zt, Greiner:1993jn, Migdal:1990vm}.

After outlining the theoretical setup in Sec. \ref{stuppa}, we present the phase diagram in the $\theta-\mu_I-\mu_s$ plane in Sec. \ref{phasediagram}. We pay special attention to the so-called Dashen's phenomenon i.e. the spontaneous breaking of CP symmetry at $\theta=\pi$ associated with a non-analytic $\theta$-dependence of the ground state energy and a phase transition as $\theta$ crosses $\pi$ \cite{Dashen:1970et, Witten:1980sp, DiVecchia:1980yfw, Smilga:1998dh, Creutz:1995wf, Creutz:2003xu, Gaiotto:2017tne, DiVecchia:2017xpu, DiVecchia:2013swa}. We start from the simpler case of degenerate quark masses, where we observe the existence of a quadruple point where the encounter between Normal (i.e. the zero density phase), Pion, and Kaon phases occurs in conjunction with the Dashen phase transition at $\theta=\pi$. Moreover, we discover that the transition between the two superfluid phases does not depend on $\theta$. Another interesting observation is the possibility of triggering the Normal to superfluid phase transitions by varying $\theta$ at fixed values of the chemical potentials. We then extend the analysis to the more physical scenario of $m_u = m_d = m \neq m_s$ unveiling an intricate phase structure with lines of phase transitions of different orders intersecting at special multi-critical points. It is worth recalling that in the Normal phase one can identify two regions, depending on the value of $\gamma\equiv m/m_s$ which are characterized by the presence/absence of spontaneous breaking of CP symmetry at $\theta = \pi$. The regions where CP is broken, corresponding to $\gamma <2$ feature the first order Dashen phase transition as the $\theta$-angle crosses $\theta = \pi$, which becomes of the second order at the critical point $\gamma=2$. One of our main results is the disappearance of the Dashen phase transition in the superfluid phases. Nevertheless, for $\gamma >2$, we discover a novel parity-preserving superfluid phase at $\theta=\pi$ characterized by the $\expval{\bar u d}$ scalar isospin order parameter rather than the pseudoscalar $\expval{\bar u \gamma_5 d}$ one associated with the Pion phase. A similar phenomenon has been recently observed in the two-flavor color superconducting phase at large baryon density \cite{Murgana:2024djt}.

We determine the boundary of the various phases, the order of the phase transitions separating them, and the properties of the multicritical lines. Our findings indicate that the phase transitions from the Normal to the superfluid phases are of second order regardless of the value of $\theta$. Similarly, the transition between the two superfluid phases remains first order even at non-vanishing $\theta$-angle. Interestingly, as in the case of degenerate quark masses, we observe the possibility of inducing a superfluid transition by varying $\theta$ while keeping $\mu_I$ and $\mu_s$ fixed. At the same time, when the masses are not degenerate, the position of the line separating the two superfluid phases depends on $\theta$. This implies that the phase transition between the two superfluid phases can be triggered solely by varying $\theta$. Notably, we find regions of the parameter space where the superfluid phases can be realized for tiny values of the corresponding chemical potentials. Of particular theoretical interest is the observation of a quadruple point where Normal, Pion, and Kaon phases meet alongside the Dashen transition at $\theta = \pi$. Intriguingly, for $\gamma =2$, the quadruple point sits exactly at $\mu_s = 0$.

We further characterize the phase diagram by determining the $\theta$-dependence of the various quark condensates and charge densities in Sec. \ref{sezioconde}. While at $\theta=0$, the positivity of the Euclidean fermion determinant leads to QCD inequalities forbidding the formation of the CP-odd $\braket{\bar u \gamma_5 u}$, $\braket{\bar d \gamma_5 d}$ \cite{Son:2000xc}, these condensates form at nonzero $\theta$ in all phases. Analogously, at finite $\theta$-angle the CP-even $\braket{\bar u d}$, $\braket{\bar u s}$ form in the Pion and Kaon phases, respectively.  The condensates exhibit a jump across the various first order phase transitions and are continuous across the second order ones. Moreover, we determine the gluon condensate whose behavior at finite density is of general phenomenological interest. 

In Sec. \ref{spectrum} we determine the spectrum of the theory in the various phases and discuss the $\theta$-dependence of the meson masses. For any value of the $\theta$-angle, each superfluid phase features a gapless mode which can be interpreted as a Goldstone boson. In the Normal phase, we identify a mode that becomes massless at the second order Dashen transition occurring at $\gamma=2$ and $\theta=\pi$.

It is useful to recall that two-color QCD at finite baryon density with degenerate quark masses displays an intriguing $\theta$-dependence strongly dependent on the number of light flavors $N_f$ being even or odd \cite{Bersini:2022jhs}. There it was observed that, for odd $N_f$, the disappearance of Dashen's phenomenon in the superfluid phase is accompanied by the emergence of two novel first order phase transitions at $\theta=\pi/2$ and $\theta=3\pi/2$. Motivated by the similarities between two-color QCD at finite $\mu_B$ and QCD at finite $\mu_I$, in Sec. \ref{generalNF} we analyze the phase diagram in the case of $N_f$ light flavors with degenerate masses for a given finite density realization. We, however, do not find any phase transition at $\theta=\pi/2$ and $\theta=3\pi/2$. Instead, we discover that, unlike the case of two and three light flavors, Dashen's phenomenon typically persists in the superfluid phases. We offer our conclusions in Sec. \ref{conclusions}.

\section{Low-energy Lagrangian}
 \label{stuppa}

The low-energy dynamics of three-flavor QCD in the presence of a nonzero $\theta$-angle can be described in terms of the chiral Lagrangian below
\begin{align}
\label{lagtheta}
    \mathcal{L} &= \frac{F_\pi^2}{4} Tr\{ \nabla_\mu\Sigma\nabla^\mu\Sigma^\dagger\}+ \frac{F_\pi^2}{2} G Tr\{M\Sigma+M^\dagger\Sigma^\dagger\}  - \frac{a F_\pi^2}{4}\left(\theta-\frac{i}{2}Tr\{\log \Sigma - \log \Sigma^\dagger \}\right)^2 \,,
\end{align}
where $F_\pi$ is the pion decay constant, $ F_\pi^2 G$ corresponds to the chiral condensate at zero $\theta$-angle, $F_\pi^2 a/2$ is the topological susceptibility of the underlying Yang-Mills theory, and the quark mass matrix is $M={\rm diag}(m_u,  m_d, m_s) $.
The matrix field $\Sigma \in U(3)$ can be written as 
\begin{equation}
\Sigma = U \Sigma_0 U \,, \qquad  U = exp \left( \frac{i \Phi}{\sqrt{2} F_{\pi}}\right) \,, \qquad \Phi = \Pi^a T^a + \frac{S}{\sqrt{3}} \,,
\end{equation}
where $\Sigma_0$ maximizes the static Lagrangian and $T^a$ are the $SU(3)$ generators normalized as $\Tr [T^a T^b ] = \delta^{ab}$. $\Sigma$ describes the pseudoscalar meson octet  
\begin{eqnarray}
\Pi^a T^a=\left( \begin{array}{ccc}
\frac{\pi^0}{\sqrt{2}}+\frac{\eta}{\sqrt{6}} & \pi^+ & K^+\\
\pi^- & -\frac{\pi^0}{\sqrt{2}}+\frac{\eta}{\sqrt{6}} & K^0 \\
K^- & \bar K^0 & -\frac{2 \eta}{\sqrt{6}}
 \end{array} \right) \,, \label{Goctet}
\end{eqnarray}
plus a singlet field $S$ related to the physical $\eta^\prime$ state. The quark chemical potential has been included in the covariant derivative, which reads
\begin{align}
\nabla_\nu \Sigma = \partial_{\nu} \Sigma- i [ B_\nu , \Sigma]  \,,
\end{align}
where
\be \label{Bmatrix}
B_\nu=  \delta_{\nu 0} B \,, \quad B= {\rm
diag} \left(\frac13 \mu_B+\frac12 \mu_I,\frac13 \mu_B-\frac12 \mu_I, \frac13
\mu_B-\mu_S \right) \,,
\ee
with $\mu_I$, $\mu_s$, and $\mu_B$ being the isospin, strangeness, and baryon chemical potentials, respectively. Since the low-energy theory describes only mesonic degrees of freedom the baryon density vanishes and $\mu_B$ does not enter the effective Lagrangian.
In what follows, we will often take the masses of up and down quarks to be equal, i.e. $m_u=m_d=m$. For vanishing chemical potentials and $\theta$-angle, the tree-level meson masses are then given by the Gell-Mann--Oakes--Renner formula
\be
  m_\pi^2=2 G m \,,\qquad
  m_K^2=  G (m+m_s)  \,, \qquad m_{\eta}^2 =\frac23 G (m + m_s) \,.
\ee

\section{Phase diagram} \label{phasediagram}

In this section, we investigate the $\theta - \mu_I - \mu_s$ phase diagram of theory. At zero $\theta$-angle the vacuum ansatz reads \cite{Kogut:2001id}
\begin{eqnarray} \label{SadP}
\Sigma_c=\left(
\begin{array}{ccc}
 \cos\varphi & i  e^{-i \xi } \cos \beta \sin \varphi & -e^{-i (\xi +\rho )}\sin \beta\sin \varphi \\
 i  e^{i \xi } \cos \beta \sin \varphi & \cos ^2 \beta \cos \varphi +\sin ^2\beta &- i e^{-i \rho } \sin (2 \beta ) \sin^2 (\varphi/2) \\
 e^{i (\xi +\rho )} \sin \beta \sin \varphi & i e^{i \rho } \sin (2 \beta ) \sin^2 (\varphi/2)& \sin ^2 \beta \cos \varphi+\cos ^2 \beta \\
\end{array}
\right)     \,.
\end{eqnarray}
As typical, we take into account the impact of the $\theta$-angle on the vacuum by introducing the Witten variables $\{\alpha_1, \alpha_2, \alpha_3 \}$ as follows \cite{Witten:1980sp}
\be
\label{eq:ansatsvacuumcomplete}
\Sigma_0 =W \Sigma_c\ , \,\, \text{with}\  \, \, W= \text{diag}\{e^{-i \alpha_1}\,,e^{-i \alpha_{2}} \,, e^{-i \alpha_{3}} \} \ .
\ee
The corresponding static Lagrangian reads
\allowdisplaybreaks
\begin{align}
    {\cal L}_0 &= \frac{1}{4} F_{\pi}^2 \Bigg(-a\bar{\theta}^2 + \mu_I^2 \sin ^2 \varphi \cos^2 \beta-\frac{1}{2} \mu_I^2 \left(\cos\varphi \cos^2\beta+\sin ^2\beta \right)^2-\frac{1}{2} \mu_I^2 \cos ^2 \varphi+\mu_I^2+2 \mu_s^2 \nonumber \\ & +2\mu_I \mu_s \sin^2\varphi \sin^2\beta-2 \mu_I \mu_s \sin ^4\left(\frac{\varphi }{2}\right) \sin ^2(2 \beta )-2 \mu_s^2 \left(\sin ^2\left(\frac{\varphi }{2}\right) \cos (2 \beta )+\cos ^2\left(\frac{\varphi}{2}\right)\right)^2 \nonumber \\ &   +4 G \left( m_u \cos \varphi \cos\alpha_1 + m_d \cos \alpha_2 \left(\cos \varphi \cos ^2 \beta+\sin ^2\beta \right) + m_s \cos \alpha_3 \left(\cos \varphi \sin^2\beta +\cos^2 \beta \right) \right)  \Bigg) \,,
\end{align}
where we introduced 
\be
 \bar \theta \equiv \theta -\alpha_1-\alpha_2-\alpha_3 \,.
\ee
The equations of motion (EOM) are
\begin{align}
 0 &=(\mu_I-2 \mu_s) \left(2 \sin (4 \beta ) (\mu_I-2 \mu_s) \sin ^4\left(\frac{\varphi }{2}\right)-\sin (2 \beta ) (3 \mu_I+2 \mu_s) \sin ^2 \varphi \right)  \nonumber\\ &+16 G \sin (2 \beta ) \sin ^2\left(\frac{\varphi }{2}\right) (m_d \cos \alpha_2-m_s \cos \alpha_3) \,, \\
    0 &= \sin (2 \varphi ) \left(4 \cos (2 \beta ) (3 \mu_I+2 \mu_s) (\mu_I-2 \mu_s)+19 \mu_I^2+20 \mu_I \mu_s+12 \mu_s^2\right)  \nonumber\\ 
 &-8 \cos (4 \beta ) (\mu_I-2 \mu_s)^2 \sin ^3\left(\frac{\varphi }{2}\right) \cos \left(\frac{\varphi }{2}\right)+2 \sin \varphi \left((\mu_I-2 \mu_s)^2 \right. \nonumber\\ 
 & \left. -32 G \left(m_d \cos \alpha_2 \cos ^2 \beta +m_s \cos \alpha_3 \sin ^2 \beta +m_u \cos \alpha_1\right)\right) \,,\\
    0 &= -a \bar \theta +2 G m_u \sin \alpha_1 \cos \varphi  \,,\\
      0 &= -a \bar \theta +2 G m_d \sin \alpha_2 \left(\cos ^2\beta  \cos \varphi +\sin ^2 \beta \right) \,,\\
        0 &= -a \bar \theta +2 G m_s \sin \alpha_3 \left(\sin ^2 \beta \cos \varphi+\cos ^2 \beta \right) \,.
\end{align}
Regardless of the solution of the last three EOM, the first two are solved by
\begin{align}
    \cos \varphi &=1  \,, & \beta \in (0,\pi/2) \,, \qquad \text{Normal phase} \,, \\ \label{cosphipionico}
        \cos \varphi  &= \frac{G (m_d \cos \alpha_2 +m_u \cos \alpha_1)}{ \mu_I^2} \,, &\beta  = 0  \,, \qquad \quad  \ \text{Pion phase} \,, \\ \label{cosphikaonico}
            \cos \varphi &= \frac{4 G (m_s \cos \alpha_3 +m_u \cos \alpha_1)}{(\mu_I+2 \mu_s)^2} \,, &\beta = \frac{\pi}{2}  \,, \qquad \ \ \text{Kaon phase} \,.
\end{align}

The angles $\xi$ and $\rho$ do not appear in the EOM and represent a residual $U(1)_I\times U(1)_s$ degeneracy of the ground state. The solution \eqref{cosphipionico} corresponds to the so-called Normal phase where $U(1)_I$ and $U(1)_s$ are unbroken and $\Sigma_0$ is the identity matrix. On the other hand, $U(1)_I$ and $U(1)_s$ are 
spontaneously broken in the superfluid Pion and Kaon phases, respectively, due to the formation of pion and kaon condensates.

The energy density in the three phases (labeled as N, P, and K) reads
\begin{align}
   E_N &=\frac{a F_\pi^2}{4}  \bar \theta^2 - F_\pi^2 G (m_u \cos \alpha_1 +m_d \cos \alpha_2 +m_s \cos \alpha_3)  \,,  \\
   E_P &=\frac{a F_\pi^2}{4}  \bar \theta^2 -\frac{F_\pi^2 }{2 \mu_I^2} \left(G^2 (m_d \cos \alpha_2+m_u \cos \alpha_1)^2+2 G \mu_I^2 m_s \cos \alpha_3+ \mu_I^4\right) \,,  \\
   E_K &=\frac{a F_\pi^2}{4}  \bar \theta^2 -\frac{F_\pi^2}{(\mu_I+2 \mu_s)^2} \left(2 G^2 (m_u \cos \alpha_1+ m_s \cos \alpha_3)^2+ G m_d (\mu_I+2 \mu_s)^2 \cos \alpha_2\right)\nonumber\\  &  -\frac{F_\pi^2}{8} (\mu_I+2 \mu_s)^2  \,.
\end{align}
 Equipped with the above results, we now analyze the impact of the $\theta$-angle on the phase diagram, starting from the case of degenerate quark masses.

\subsection{$m_u = m_d = m_s = m$} \label{masseuguali}

As a warm-up analysis, we determine the phase diagram for degenerate quarks. To this end, we consider the $a \gg m_\pi$ limit equivalent to decoupling the singlet mode from the spectrum and directly incorporating the $\theta$-angle into the quark mass matrix. As a result, the Witten variables are constrained to satisfy $\bar \theta = 0 \ \text{Mod 2} \pi$ \footnote{The modulo stems from the fact that when a solution $\{\alpha_1, \alpha_2, \alpha_3 \}$ of the EOM is found, then it becomes possible to construct another solution as follows
\be
\alpha_1(\theta+2\pi)=\alpha_1(\theta)+ 2 \pi \,, \qquad  \alpha_2(\theta+2\pi)=\alpha_2(\theta) \,, \quad \alpha_3(\theta+2\pi)=\alpha_3(\theta) \,.
\ee
However, since $\Sigma$ depends only on $e^{-i \alpha_i}$  ($i=1,2,3$), the physics is invariant under $\theta \to \theta +2 \pi$.}. In the Normal phase, we have $\varphi=0$ and the remaining EOM reduce to
\begin{align}
    \sin \alpha_i =\sin \alpha_j \,, \quad i,j=1,2,3 \,.
\end{align}
The energy is minimized when $\alpha_1=\alpha_2=\alpha_3=\alpha(\theta)$ with $\alpha(\theta)$ given by
\be \label{questaequazione3}
\alpha(\theta)=\begin{cases}
    \frac{\theta}{3}  \,, \qquad  \ \ \theta \in [0, \pi] \,, \\
     \frac{\theta -2\pi}{3}\,,  \ \ \ \ \theta \in [\pi, 2\pi]  \,.
\end{cases} 
\ee

At $\theta=\pi$ the two solutions for $\alpha(\theta)$ become degenerate and related by a CP transformation that acts as $\Sigma \to \Sigma^\dagger$. Therefore, the Normal phase exhibits spontaneous breaking of CP symmetry at $\theta=\pi$, and, as $\theta$ crosses $\pi$, a first order phase transition occurs. The ground state energy reads
\be \label{normenergy}
E_N/ F_\pi^2= - \frac{3}{2}m_\pi^2 \cos(\alpha(\theta))\,.
\ee
\vskip 1em
Next, we consider the Pion phase corresponding to $\beta=0$ and 
\be
  \cos \varphi  = \frac{m_\pi^2}{2\mu_I^2} ( \cos \alpha_2+ \cos \alpha_1) \,. 
\ee
By plugging the above into the EOM for the Witten variables, we arrive at
\begin{align}
     \alpha_1+\alpha_2+\alpha_3 =\theta \,, \qquad \sin \alpha_1 &=\sin  \alpha_2 \,, \nonumber\\
     \frac{ m_\pi^2 }{2 \mu_I^2} \sin \alpha_1 (\cos \alpha_1+\cos \alpha_2)&=\sin \alpha_3 \,.   
\end{align}
One can either have $\alpha_1=\alpha_2$ or $\alpha_1=\pi-\alpha_2$. The energy is minimized when considering the former solution which leads to the following equations
\begin{align} \label{equalpion}
    2 \alpha_1 =\theta-\alpha_3 \,, \qquad
    \frac{m_\pi^2 }{2 \mu_I^2} \sin (\theta -\alpha_3)=\sin \alpha_3 \,,
\end{align}
whose solutions read
\be
\cos \alpha_3^{(\pm)} =\pm\frac{m_\pi^2 \cos \theta +2 \mu_I^2}{\sqrt{m_\pi^4+4  \mu_I^2  m_\pi^2 \cos \theta +4 \mu_I^4}}\ .
\ee
The ground state of the theory corresponds to the $\alpha_3^{(+)}$ solution which yields 
\begin{align}
E_P/F_\pi^2= -\frac{2 \mu_I^4+m_\pi^4 + m_\pi^2\sqrt{4 \mu_I^4+m_\pi^4+4 \mu_I^2 m_\pi^2 \cos \theta}}{4 \mu_I^2} \,.
\end{align}
For $\theta =0$, the above reduces to
\be
E_P/F_\pi^2 \rvert_{\theta=0} = -\frac{1}{2\mu_I^2} \left(m_\pi^4+m_\pi^2 \mu_I^2 +\mu_I^4\right) \,,
\ee
in agreement with \cite{Kogut:2001id}. 
Unlike the Normal phase, the physics is now analytic in $\theta$ everywhere, and the spontaneous breaking of $CP$ at $\theta=\pi$ is absent. The transition from the Normal to the Pion phase is of the second order and occurs at a critical value of $\mu_I$ given by
\be
\mu^2_{I,cr}(\theta) = m_\pi ^2 \cos \left(\alpha(\theta)\right) \,,
\ee
with $\alpha(\theta)$ in Eq. \eqref{questaequazione3}. The minimum of $\mu_{I,cr}(\theta)$ occurs at $\theta=\pi$ with $\mu_{I,cr}(\theta=\pi) = \frac{m_\pi}{\sqrt{2}}$. 

\begin{figure}[t!]
\centering
 \includegraphics[width=0.69\textwidth]{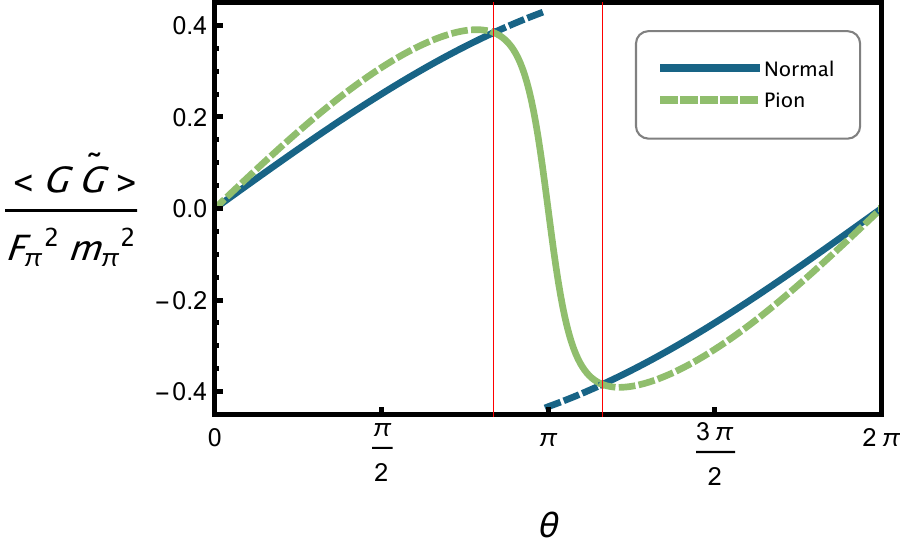} 
	\caption{The normalized CP order parameter $\braket{G \Tilde{G}} = - \frac{\partial E}{\partial \theta}$ as a function of $\theta$ for $\mu_I =\frac{4}{5}m_\pi$. The blue and green lines have been obtained by computing $\braket{G \Tilde{G}}$ according to the expressions of the energy in the Normal and Pion phases, respectively. The true behavior corresponds to the solid portion of these lines. The red vertical line marks the second-order superfluid phase transition, across which $\braket{G \Tilde{G}}$ is continuous.} 
	\label{CPchange}
\end{figure}

Finally, in the Kaon phase we obtain the same EOM of the Pion phase upon exchanging $\alpha_2$ and $\alpha_3$ and replacing $\mu_I^2 \to \frac{(\mu_I+2 \mu_s)^2}{4}$. Therefore, we can readily determine the ground state energy 
\begin{align}
    E_K/F_\pi^2 &= -\frac{1}{8} (\mu_I+2 \mu_s)^2-\frac{2m_\pi^4 - m_\pi^2 \sqrt{(\mu_I+2 \mu_s)^4+4 m_\pi^4+4 m_\pi^2 (\mu_I+2 \mu_s)^2 \cos \theta }}{2 (\mu_I+2 \mu_s)^2}\,.
\end{align}
which for $\theta=0$ simplifies to
\be
E_K/F_\pi^2 \rvert_{\theta=0} =-\frac{1}{8} (\mu_I+2 \mu_s)^2-\frac{2 m_\pi^4}{(\mu_I+2 \mu_s)^2}-\frac{m_\pi^2}{2} \,,
\ee
in agreement with \cite{Kogut:2001id}.  The critical value of $\mu_s$ that marks the second order phase transition from the Normal to the Kaon phase is
\be \label{moshimoshi}
 \mu_{s,cr}(\theta)= m_\pi \sqrt{ \cos \left(\alpha(\theta)\right)} - \frac{\mu_I}{2}\ .
\ee
The dependence of both $\mu_{s,cr}$ and $\mu_{I,cr}$ from $\theta$ opens the intriguing possibility of triggering the superfluid phase transition by varying $\theta$ at fixed values of the chemical potential. To illustrate this phenomenon, we show the normalized CP order parameter $\braket{G \Tilde{G}} = -\frac{\partial E}{\partial \theta}$ as a function of $\theta$ in Fig. \ref{CPchange} for $\mu_s=0$ and $\mu_I =\frac45 m_\pi$. One can note the occurrence of two superfluid phase transitions at $\theta= 3 \cos ^{-1}\left(\frac{16}{25}\right) $ and $\theta=2 \pi - 3 \cos ^{-1}\left(\frac{16}{25}\right) $. Moreover, it is evident that Dashen's phenomenon disappears in the Pion phase.
To fully unveil the phase diagram of the theory, we need to determine the curve separating the Pion and Kaon phases. Interestingly, the latter shows no dependence on $\theta$. In fact, by comparing the energies in the two phases, we conclude that the Kaon phase is realized when
\be
\mu_s > \mu_I/2 \,, \qquad \mu_s >  \mu_{s,cr}(\theta)\ ,
\ee
with the transition between the two superfluid phases being of the second order. As we shall see in the next section, the transition between the Pion and Kaon phases does not depend on the $\theta$ angle only in the limit of degenerate quark masses. 

Finally, by combining the above with Eq. \eqref{moshimoshi} we obtain that Normal, Pion, and Kaon phases meet at the special tricritical point defined by
\be
\mu_s = \frac{\mu_I}{2} = \frac{m_\pi}{2} \sqrt{\cos(\alpha(\theta))} \,.
\ee
Intriguingly, at $\theta=\pi$ the above reduces to $\mu_s = \frac{\mu_I}{2} =\frac{m_\pi}{2 \sqrt{2}}$, which corresponds to a quadruple point where the Normal, Pion, and Kaon phases meet the Dashen phase transition.

\subsection{$m_u = m_d = m \neq m_s$}

\begin{figure}[t!]
\centering
\includegraphics[width=0.49\textwidth]{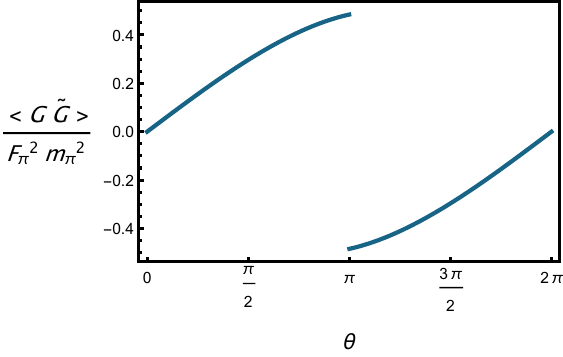} \includegraphics[width=0.49\textwidth]{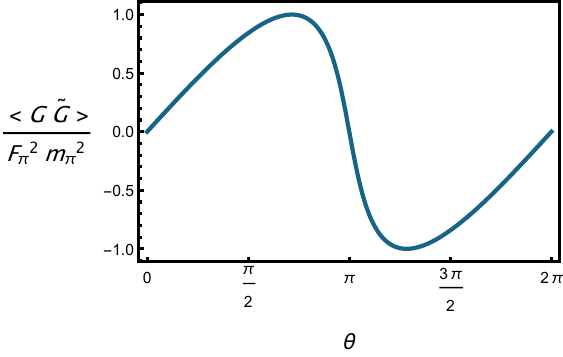} 
	\caption{The normalized CP order parameter $\braket{G \Tilde{G}}$ in the Normal phase for $\gamma=1/2$ (\textit{left}) and $\gamma=3$ (\textit{right}).} 
	\label{CPfasenorm}
\end{figure}

We now move to the more complicated case where only the up and down masses are considered equal and we consider the limit $a \gg m_\pi^2, m_K^2$. The EOM in the Normal phase read
\begin{align}\label{norm1}
    \alpha_1+\alpha_2+\alpha_3 = \theta  \,,  \quad  \sin \alpha_1 = \sin \alpha_2 \,, \quad m \sin \alpha_1= m_s \sin \alpha_3\ .
\end{align}
The minimum of the energy is achieved when the Witten variables associated with the up and down quarks are equal, i.e.
\be \label{baralfa}
\alpha_1=\alpha_2 \equiv \alpha = \frac{\theta-\alpha_3}{2} \,.
\ee
The relevant solutions for $\alpha_3$ read
{\small
\begin{align} \label{solnormal}
    \alpha_3^{(\pm)} &= 2 \cos ^{-1}\left(\frac{1}{12} \left(-3 \gamma  \cos \left(\frac{\theta }{2}\right)\pm \sqrt{\frac{9}{2} \gamma ^2 (\cos \theta+1)+\frac{3 \left(-\gamma ^2+x+4\right)^2}{x}} \right. \right. \nonumber \\ & \left. \left.  \pm \sqrt{9 \gamma ^2 (\cos \theta+1)-12 \left(\gamma ^2-4\right)\mp \frac{9 \gamma  \cos \left(\frac{\theta }{2}\right) \left(\gamma ^2 \cos \theta-\gamma ^2-8\right)}{\sqrt{\frac{1}{2} \gamma ^2 (\cos \theta+1)+\frac{\left(-\gamma ^2+x+4\right)^2}{3 x}}}-\frac{3 \left(\gamma ^2-4\right)^2}{x}-3 x}\right)\right) \,,  \nonumber \\ x & \equiv \Big(3 \sqrt{6} \gamma^2\sqrt{ \sin^2 \theta  \left(2 \gamma ^6-27 \gamma ^4 \cos (2 \theta )+3 \gamma ^4+96 \gamma ^2-128\right)} \nonumber \\ &+ \gamma ^6+ \gamma ^4(15-27 \cos (2 \theta )) +48 \gamma ^2-64\Big)^{1/3} \,,
\end{align}}
where we introduced the quark mass ratio
\begin{equation}\label{mgammams}
 \gamma \equiv m/m_s = \frac{m_\pi^2}{2 m_K^2-m_\pi^2} \,.
\end{equation}
For $\gamma \le 2$, the solutions $\alpha_3^{(+)}$ and  $\alpha_3^{(-)}$ minimize the energy in the intervals $\theta \in [0,\pi]$ and $\theta \in [\pi, 2\pi]$, respectively. At $\theta=\pi$ the two solutions become degenerate and related by CP transformations. As $\theta$ crosses $\pi$, the theory undergoes a first order phase transition and the Dashen's phenomenon occurs. This can be better seen by considering the limit of a large strange quark mass, namely $\gamma \ll 1$. In fact, the two solutions above can be expanded as
\be
  \alpha_3^{(\pm)}  = \pm  \sin \left(\frac{\theta}{2}\right)  \gamma-\frac{1}{4} \sin \theta  \ \gamma ^2 + \mathcal{O}\left(\gamma^3 \right) \,,
\ee
with the corresponding values of the energy being
\begin{align}
    E_N^{(\pm)} =F_\pi^2(m_K^2-m_\pi^2/2) \left(-1 \mp 2 \cos(\theta/2) \gamma + \frac{1}{4} (\cos \theta -1) \gamma ^2 + \mathcal{O}\left(\gamma^3 \right) \right) \,.
\end{align}
At higher orders, the even powers of $\gamma$ in the energy are the same for both solutions, while the odd powers of $\gamma$ appear with opposite coefficients which, however, vanish for $\theta = \pi$.
\vskip .5em
For $\gamma >2$, the theory enters a region with unbroken CP symmetry at $\theta=\pi$. In fact, while for $\gamma<2$ there are two degenerate solutions at $\theta=\pi$, namely
\be
\alpha_3^{(\pm)}\rvert_{\theta=\pi} = 2 \sec ^{-1}\left(\pm\frac{2 \sqrt{2}}{\sqrt{4-\gamma ^2}}\right) \,,
\ee
for $\gamma \ge 2$, the ground state at $\theta=\pi$ is the CP preserving solution $\alpha_3 = \pi \,,  \alpha = 0$. While Eq. \eqref{solnormal} still corresponds to the ground state of the theory, the solutions $\alpha_3^{(\pm)}$ can now be glued together at $\theta=\pi$ and the resulting ground state energy becomes an analytic function of $\theta$. To illustrate this fact, we show the CP order parameter $\braket{G \Tilde{G}}$ in Fig. \ref{CPfasenorm} for $\gamma=1/2$ and $\gamma=3$. The value $\gamma=2$ corresponds to the critical point at the end of the first order line. At this critical point, the topological susceptibility $\chi \propto \frac{\partial^2 E}{\partial \theta^2}$ diverges at $\theta=\pi$ signaling that the phase transition becomes of the second order. 

\vskip 1em
In the Pion phase, one still has Eq. \eqref{baralfa} but $\alpha_3$ now satisfies
\begin{align}
  \frac{m_\pi^4}{2 \mu_I^2 (2 m_k^2 - m_\pi^2)}  \sin (\theta -\alpha_3)=\sin \alpha_3 \,.
\end{align}
The above generalizes Eq. \eqref{equalpion} and yields
\be \label{alpha3pion}
\cos \alpha_3^{(\pm)} =\pm\frac{2 \mu_I^2 \left(2 m_K^2-m_\pi^2\right)+m_\pi^4 \cos \theta}{\sqrt{4 \mu_I^4 \left(2 m_K^2- m_\pi^2\right)^2+4 \mu_I^2 m_\pi^4 \left(2 m_K^2-m_\pi^2\right) \cos \theta +m_\pi^8}} \,,
\ee
with the lowest energy solution being $\alpha_3^{(+)}$.
The corresponding ground state energy reads
\begin{align}
    E_P/F_\pi^2 = -\frac{2 \mu_I^4+ m_\pi^4+\sqrt{m_\pi^8 +4 \mu_I^4 (m_\pi^2-2 m_K^2)^2+ 4 \mu_I^2 m_\pi^4 (2 m_K^2-m_\pi^2) \cos \theta }}{4 \mu_I^2} \,.
\end{align}

\begin{figure}[t!]
\centering
 \includegraphics[scale=0.7]{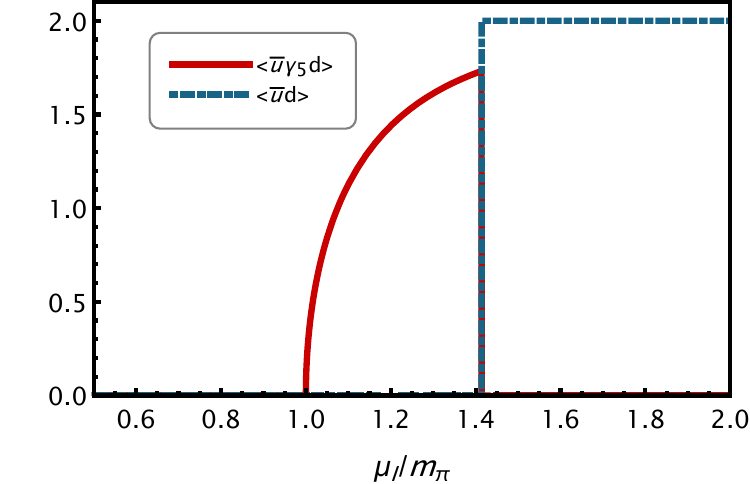} 
	\caption{Scalar $\langle \bar u d \rangle$ and pseudoscalar $\langle \bar u \gamma_5 d \rangle$ condensates in units of $F_\pi^2 G$ as a function of $\mu_I/m_\pi$ at $\theta=\pi$ and $\gamma = 4$. One can observe three distinct regions. First, for $\mu_I < m_\pi$ the theory is in the Normal phase and both condensates are zero. Second, for $m_\pi <\mu_I < \mu_I^* = \sqrt{2} m_\pi$ the theory is in the Pion phase with broken parity. Finally, for $\mu_I > \mu_I^*$ the theory is in a different superfluid phase where parity is restored.} 
	\label{condensa}
\end{figure}

\begin{figure}[t!]
    \centering
    \includegraphics[scale=0.8]{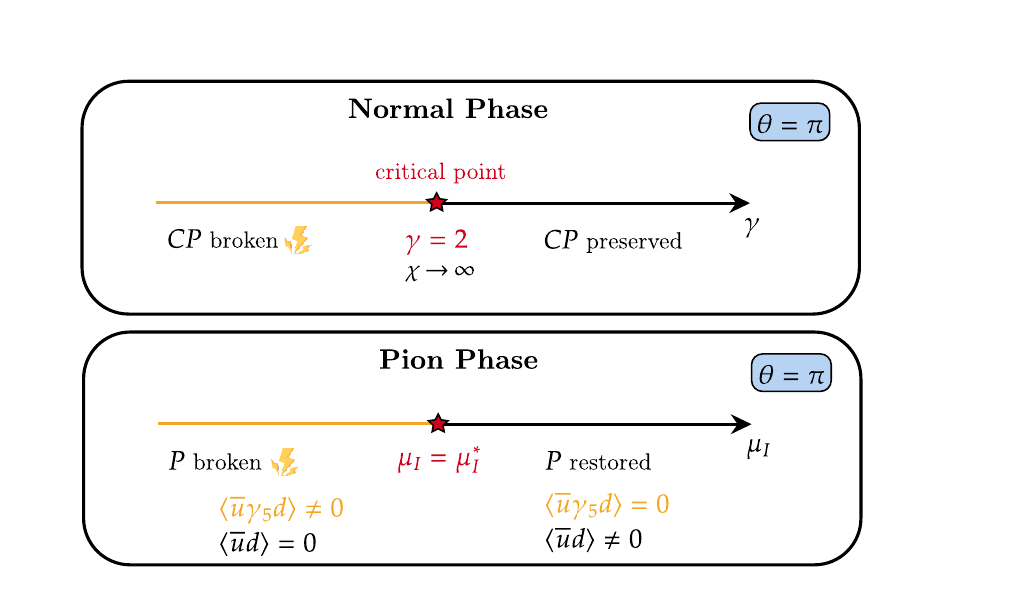}
    \caption{{\it Top panel}: the fate of $CP$ symmetry at $\theta = \pi$ in the Normal phase. {\it Bottom panel}: the fate of $P$ symmetry at $\theta = \pi$ in the Pion phase ($\mu_I < \mu_I^*$) and the superfluid phase associated with the $\langle \bar u d \rangle$ condensate ($\mu_I > \mu_I^*$).}
    \label{fig:summaryNP}
\end{figure}

% \begin{align}
%     E_P/F_\pi^2 =-\frac{\mu_I^2}{2}+\frac12\left(m_\pi^2-2 m_K^2\right) \cos \left(2 \cot ^{-1} A_P\right)-\frac{m_\pi^4 \cos ^2\left(\frac{\theta }{2}+\cot ^{-1} A_P\right)}{2\mu_I^2} \,,
% \end{align}
% where
% \begin{align}
%     A_P =\frac{m_\pi^4 \sin \theta}{m_\pi^4 \cos \theta- \left(2 m_K^2-m_\pi^2\right) \left(\sqrt{\frac{16 \mu_I^4 m_K^4-16 \mu_I^4 m_K^2 m_\pi^2-4 \mu_I^2 m_\pi^4 \left(m_\pi^2-2 m_K^2\right) \cos \theta +m_\pi^8+4 \mu_I^4 m_\pi^4}{ \left(m_\pi^2-2 m_K^2\right)^2}}-2\mu_I^2\right)} \,.
% \end{align}
In the $\gamma \ll 1$ limit, the energy reduces to
\be
 E_P/F_\pi^2 =-\frac{m_\pi^2}{2}\frac{1}{\gamma}-\frac{2 \mu_I^4+m_\pi^4 \cos \theta+m_\pi^4}{4 \mu_I^2}-\frac{m_\pi^6 \sin^2 \theta }{16 \mu_I^4}\gamma  + \cO\left(\gamma^2\right)\ .
\ee
For generic values of $\mu_I$ the energy of the Pion phase is an analytic function of $\theta$ and Dashen's crossover is absent. However, the $\theta$-dependence of the vacuum changes dramatically at the specific parameter point 

\be \label{50special}
\mu_{I}^* = m_\pi \sqrt{\frac{\gamma}{2}} = \frac{m_\pi^2}{\sqrt{2 \left(2 m_K^2-m_\pi^2\right)}} \,.
\ee
In fact, at $\theta=\pi$ the solution Eq. \eqref{alpha3pion} can be written as
\begin{equation}
    \cos\alpha_3^{\pm}=\frac{\mu_I^2-\mu_I^{\ast 2}}{\abs{\mu_I^2-\mu_I^{\ast 2}}} \,,
\end{equation}
and so $\mu_I = \mu_I^\ast$ plays a role analogous to $\gamma=2$ in the Normal phase. In particular, for $\mu_I < \mu_{I}^*$, the vacuum at $\theta=\pi$ is unique and given by $\alpha_3 = \pi$, $\alpha = 0$, whereas for $\mu_I > \mu_{I}^*$ there exist two degenerate minima given by $\alpha_3 = 0$, $\alpha =\pm \pi/2$. Despite this, the CP order parameter remains zero at $\theta = \pi$ for any $\mu_I > \mu_{I}^*$. This apparent paradox is resolved by observing that at the special parameter point $\mu_I =\mu_{I}^*$, $\theta=\pi$ a first order phase transition to a different superfluid phase takes place. As shown in Fig. \ref{condensa}, the corresponding order parameter is the scalar condensate $\langle \bar u d \rangle$ which replaces the ordinary pseudoscalar order parameter $\langle \bar u \gamma_5 d  \rangle$ associated with the Pion phase at $\theta=0$. Consequently, this transition is associated with the restoration of parity symmetry for $\mu_I > \mu_{I}^*$. The analytic expression of the quark condensates will be provided in Sec. \ref{condensa}. At the special point $\mu_I = \mu_{I}^*$ the energy reduces to
\be
 E_P/F_\pi^2 =\frac{1}{4} \left(4 \left(m_\pi^2-2 m_K^2\right) \left| \cos \left(\frac{\theta }{2}\right)\right| +\frac{m_\pi^4}{m_\pi^2-2 m_K^2}-4 m_K^2+2 m_\pi^2\right) \,,
\ee
and Dashen's phenomenon occurs. Before delving into the study of the superfluid transition from the Normal to the Pion phase, we summarize the realization of the discrete symmetries at $\theta = \pi$ in Fig. \ref{fig:summaryNP}. 
%\FloatBarrier
%\FloatBarrier
% \FloatBarrier
% \FloatBarrier

The $\theta$-dependence of the critical value of $\mu_I$ for the onset of the Pion phase displays interesting features depending on the ratio of the quark masses $\gamma$. In general, we have
\be \label{mucrgen}
\mu_{I,cr}(\theta) = m_\pi \sqrt{ \cos \left( \frac{\theta-\alpha_3^{(\pm)}}{2} \right)} \,,
\ee
where $\alpha_3^{(+)}$ and $\alpha_3^{(-)}$ correspond to the regions before and after $\theta = \pi$, respectively, as discussed earlier. In the limit where the strange quark decouples from the dynamics ($\gamma \ll 1$) the above simplifies to
\be \label{tino}
 \mu_{I,cr} (\theta) = m_\pi \sqrt{\cos \left(\frac{\theta}{2}\right)}  \,,
\ee
which vanishes at $\theta=\pi$. This behavior can be traced to a vanishing leading order mass term in the $N_f=2$ chiral Lagrangian at $\theta=\pi$. Nevertheless, as first noted in \cite{Smilga:1998dh}, one needs to take into account higher-order mass terms which are expected to induce a tiny but nonzero value for $\mu_{I,cr}(\theta=\pi)$. 

\begin{figure}[t!]
    \centering
    \includegraphics{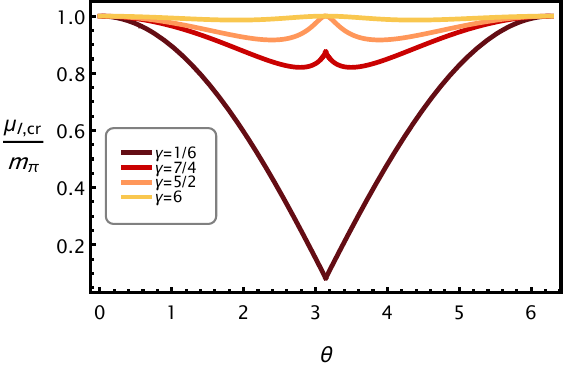}
    \caption{Critical isospin chemical potential as a function of $\theta$ for various values of the quark mass ratio $\gamma$.}
    \label{mucriticoNP}
\end{figure}

\begin{figure}[t!]
    \centering
    \includegraphics[scale=0.8]{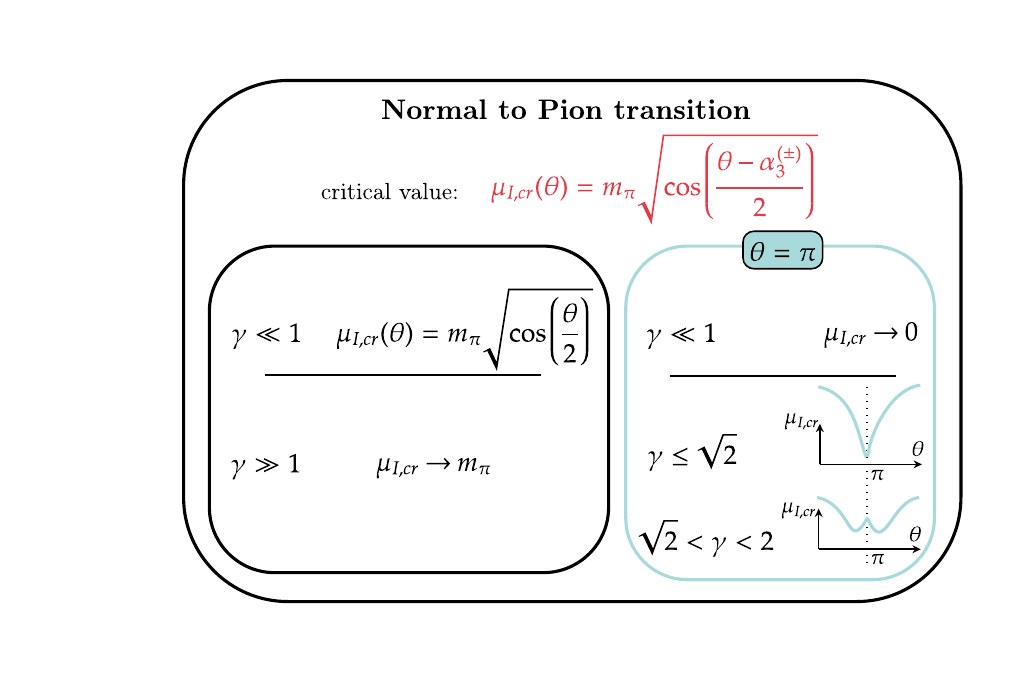}
    \caption{Summary of the $\theta$-dependence of the critical value of the isospin chemical potential separating the Normal and Pion phases.}
    \label{fig:transnormals}
\end{figure}

For $\gamma \le \sqrt{2}$ the minimum of $\mu_{I, cr}(\theta)$ occurs at $\theta=\pi$ where we have $\mu_{I,cr}(\theta =\pi) = \mu_{I}^*$. Interestingly, for $\sqrt{2}< \gamma < 2$ we still have $\mu_{I,cr}(\theta =\pi) = \mu_{I}^*$ but this value is no longer a minimum of $\mu_{I,cr}(\theta)$ but a local maximum. For $\gamma \ge 2$ the SSB of CP symmetry at $\theta=\pi$ is absent in the Normal phase and this rearrangement of the vacuum is reflected by the fact that $\mu_{I,cr}(\theta=\pi)$ becomes independent of $\gamma$ and is equal to $m_\pi$. Thus, the parity-restoring transition at $\mu_I = \mu_{I}^*$ can only take place for $\gamma >2$. Finally, for large values of $\gamma$, the $\theta$-dependence becomes increasingly softer in agreement with no physical effects of the $\theta$-angle for $m_s= 0$.  The described behavior of $\mu_{I,cr}(\theta)$ for various values of the quark mass ratio is displayed in Fig. \ref{mucriticoNP}. Despite the interesting $\theta$-dependence of $\mu_{I,cr}$ the superfluid phase transition is a second order one for any $\theta$ and $\gamma$.
The $\theta$-dependence of $ \mu_{I,cr}$ is summarized in Fig. \ref{fig:transnormals}.
% \FloatBarrier
% \FloatBarrier

\begin{figure}[t!]
\centering
\includegraphics[scale=0.75]{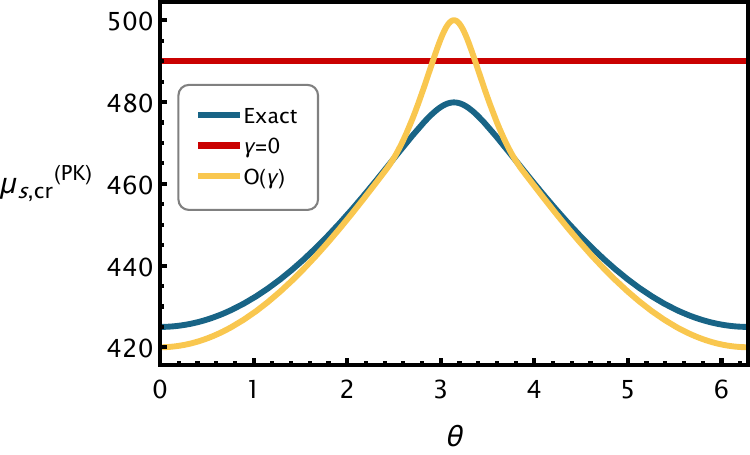} 
	\caption{$\mu_{s,cr}^{(PK)}$ (in MeV) as a function of $\theta$ ({\it Blue}) and its expansion around $\gamma=0$ given in Eq. \eqref{muspert} to the leading ({\it Red}) and next-to-leading order ({\it Yellow}) for $m_\pi = \mu_I = 140 $ MeV and $m_K = 495 $ MeV for which $\gamma \sim 0.04$.} 
	\label{cumpa}
\end{figure}

\begin{figure}[t!]
\centering
\includegraphics[scale=0.75]{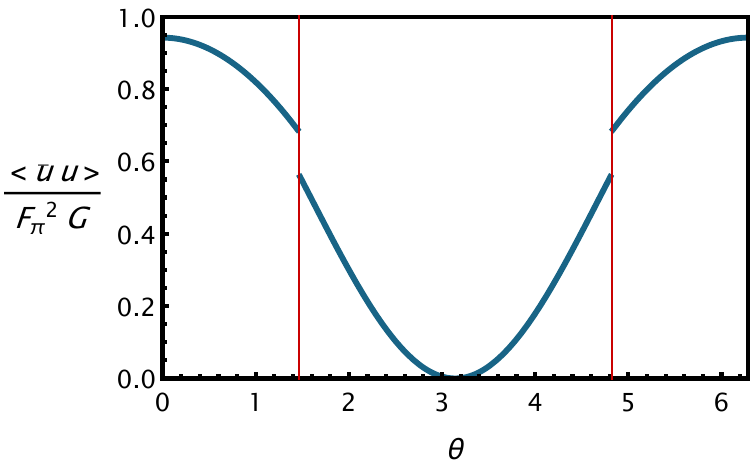}
	\caption{$\expval{\bar u u}$ as a function of the $\theta$-angle at fixed values of the chemical potentials $\mu_I =140 $ MeV and $\mu_s = 440 $ MeV. The considered values of the meson masses are $m_\pi = 140 $ MeV and $m_K =495 $ MeV. The vertical red line marks the $\theta$-induced first order transitions separating the two superfluid phases.} 
	\label{thetachange}
\end{figure}

Lastly, the EOM in the Kaon phase are
\begin{align}
\alpha_1 + \alpha_2 +\alpha_3 =\theta \,, \qquad
  \gamma \sin \alpha_1 &= \sin \alpha_3 \,, \nonumber \\
  \frac{4 m_K^2}{\gamma \,
 (1+\gamma) (\mu_I+2 \mu_s)^2} \sin \alpha_3 (\gamma  \cos \alpha_1+\cos \alpha_3) & = \sin (\theta -\alpha_1-\alpha_3) \,.
\end{align}
The solution minimizing the potential has a rather cumbersome expression which reads
\begin{align} \label{kaonsol}
    \cos \alpha_3 = \bar A_K  \sqrt{A_K} \,, \qquad
    \cos \alpha_1 = -  \sqrt{A_K/2} \,,
    \end{align}
    where
    \begin{align}
    A_K &= \frac{1}{\left(J^2+2 J \cos \theta+1\right) \left(\gamma ^4-2 \gamma ^2 \cos (2 \theta )+\left(\gamma ^2-1\right)^2 J^2+2 \left(\gamma ^2-1\right)^2 J \cos \theta+1\right)} \nonumber \\ & \times \bigg( -\gamma ^2 \left(4 J^4+\left(16 J^2+13\right) J \cos \theta+\left(11 J^2+3\right) \cos (2 \theta )+13 J^2+3 J \cos (3 \theta )+1\right)   \nonumber \\ &+ \gamma ^3 (\cos \theta-\cos (3 \theta )-2 J \cos (2 \theta )+2 J) \sqrt{J^2+2 J \cos \theta+1} \nonumber \\ & + 2 \gamma ^4 (\cos \theta+J)^2 \left(J^2+2 J \cos \theta+1\right)+2 \left(J^2+2 J \cos \theta+1\right)^2\bigg) \,, \\
    \bar A_K &=  \frac{\sqrt{2} \left(\gamma  \sin^2 \theta -\left(\gamma ^2-1\right) (\cos \theta+J) \sqrt{J^2+2 J \cos \theta+1}\right)}{\gamma ^2 \cos (2 \theta )+\gamma ^2+2 \left(\gamma ^2-1\right) J^2+4 \left(\gamma ^2-1\right) J \cos \theta-2}  \,, \\
    J &\equiv \frac{8 m_K^4}{(1+\gamma)^2(\mu_I+2 \mu_s)^2}   \,.
\end{align}
The corresponding ground state energy is given by
% \begin{align*}
%     E_K/F_\pi^2 & = -\frac{(\mu_I+2 \mu_s)^2}{8} + \frac{ A_K \left(2 \bar A_K  \gamma  \left(\sqrt{2}-\bar A_K  \gamma \right)-1\right) \left(m_\pi^2-2 m_K^2\right)^2}{4(\mu_I+2 \mu_s)^2} \nonumber \\ & +\frac{1}{2} m_\pi^2 \cos \left(\text{sgn}(\pi -\theta ) \left(\sin ^{-1}\left(\frac{\sqrt{A_K}}{\sqrt{2}}\right)-\sin ^{-1}\left(\sqrt{A_K} \bar A_K \right)\right)+\theta \right) \,,
% \end{align*}
\begin{align}
    E_K/F_\pi^2 & =  \frac{ m_K^2 \gamma}{1+\gamma} \cos \left(\text{sgn}(\pi -\theta ) \left(\sin ^{-1}\left(\frac{\sqrt{A_K}}{\sqrt{2}}\right)-\sin ^{-1}\left(\sqrt{A_K} \bar A_K \right)\right)+\theta \right) \nonumber \\ &  -\frac{(\mu_I+2 \mu_s)^2}{8} + \frac{ m_K^2 A_K}{(\mu_I+2 \mu_s)^2 (1+\gamma)^2} \left(2 \bar A_K  \gamma  \left(\sqrt{2}-\bar A_K  \gamma \right)-1\right) \,,
\end{align}
and, despite the appearances, is an analytic function of $\theta$ \footnote{This can be seen by noting that the factor multiplying the sgn function vanishes for $\theta=\pi$.}. To better illustrate this point, we expand the above around $\gamma = 0$
\begin{align}
      E_K/F_\pi^2 & = -\frac{m_\pi^4}{2  (\mu_I+2 \mu_s)^2}\frac{1}{\gamma ^2}-\frac{m_\pi^4}{(\mu_I+2 \mu_s)^2}\frac{1}{\gamma}-\frac{1}{8} \left(\frac{(\mu_I+2 \mu_s)^4+4 m_\pi^4}{(\mu_I+2 \mu_s)^2}+4 m_\pi^2 \cos \theta\right) \nonumber \\ &-\frac{1}{8} \sin^2 \theta  (\mu_I+2 \mu_s)^2  \gamma -\frac{1}{16}  \sin^2 \theta  (\mu_I+2 \mu_s)^2 \left(\frac{\cos \theta (\mu_I+2 \mu_s)^2}{m_\pi^2}+4\right)\gamma ^2 + \mathcal{O}\left( \gamma^3\right) \,.
\end{align}
The transition between the Normal and Kaon phases is of the second order and occurs at the following critical value of the strangeness chemical potential
\begin{equation}
    \mu_{s,cr}^{(NK)}(\theta) = \sqrt{m_K^2 \cos \alpha_3^{(\pm)}+\frac{1}{2}  m_\pi^2 \left(\cos \left(\frac{\theta -\alpha_3^{(\pm)}}{2}\right)-\cos \alpha_3^{(\pm)}\right)}-\frac{\mu_I}{2} \,,
\end{equation}
with $\alpha_3^{(\pm)}$ given in Eq. \eqref{solnormal}. For $\gamma < 2$ the vacuum of the Normal phase breaks spontaneously CP and accordingly $ \mu_{s,cr}^{(NK)}(\theta) $ is not analytic in $\theta=\pi$. In particular, in the $\gamma \ll 1$ limit, we obtain
\be
 \mu_{s,cr}^{(NK)}(\theta) = m_K-\frac{\mu_I}{2} + \frac{m_K}{2} \left(\left| \cos \left(\frac{\theta }{2}\right)\right| -1\right) \gamma   + \cO \left(\gamma^2 \right) \,.
\ee
Moreover, for $\gamma \le 2$ the critical $\mu_s$ attains its minimum value in $\theta=\pi$ where we have
\be \label{pi1} 
\mu_{s,cr}^{(NK)}(\theta = \pi) = \sqrt{m_K^2-\frac{m_\pi^2}{2}}-\frac{\mu_I}{2}  \,.
\ee
On the other hand, for $\gamma > 2$,  $\mu_{s,cr}^{(NK)}(\theta)$ becomes an analytic function of $\theta$. Its minimum is again achieved in $\theta=\pi$ but with the minimal value being
\be 
\label{pi2} \mu_{s,cr}^{(NK)}(\theta = \pi) = \sqrt{m_\pi^2- m_K^2}-\frac{\mu_I}{2}  \,.
\ee
By comparing Eq. \eqref{pi1} and Eq. \eqref{pi2} we see that the minimum possible value of $ \mu_{s,cr}^{(NK)}(\theta)$ is attained for $\theta=\pi$ and $\gamma = 2$ and reads 
\be \label{2theta}
 \mu_{s,cr}^{(NK)}(\theta = \pi) \rvert_{\gamma =2} = \frac{m_K}{\sqrt{3}}-\frac{\mu_I}{2}  \,.
\ee
Equation \eqref{2theta} together with  $\mu_{I,cr}(\theta=\pi) = m_\pi$ implies a curious feature of the phase diagram. In fact, for $\gamma=2$ and $\theta=\pi$, the three phases meet exactly at $\mu_s=0$, and, therefore, the Kaon phase can be realized for tiny values of $\mu_s$. In general, the location of the meeting point and the value of the critical strangeness chemical potential $ \mu_{s,cr}^{(PK)}$ separating the two superfluid phases have to be determined numerically. In the $\gamma \ll 1$ limit (at fixed $m_s$), we obtain
% {\footnotesize \begin{equation} 
%     \mu_{s,cr}^{(PK)}=\frac{1}{2} \sqrt{\mu_I^2+4 m_K^2}-\frac{  \left(\mu_I+\sqrt{\mu_I^2+4 m_K^2}\right)^3 \left(\sin ^2\theta  \left(\mu_I \left(\mu_I+\sqrt{\mu_I^2+4 m_K^2}\right)+2 m_K^2\right)+4 m_K^2 (1+\cos \theta )\right)}{16\mu_I \left(2  m_K^2 \left(2 \mu_I+\sqrt{\mu_I^2+4 m_K^2}\right)+\mu_I^3+  \mu_I\sqrt{\mu_I^2+4 m_K^2}\right)} \gamma+\order{\gamma^2}\,,
%     \label{muspert}
% \end{equation}}
\begin{equation}\small
    \mu_{s,cr}^{(PK)}=\frac{c_2}{2} -\frac{8   \left(2 m_K^2-m_{\pi }^2\right) \left(\left(c_2+\mu _I\right)^2 \sin \left(\sec ^{-1}\left(\sqrt{2c_1}\csc (\theta )\right)+\theta \right)+2 \sqrt{1-\frac{1}{2 c_1}} \left(2 m_K^2-m_{\pi }^2\right)\right)}{32 \mu _I \left[\mu _I \left(c_2+\mu _I\right)+4 m_K^2-2 m_{\pi }^2\right]}\gamma+\order{\gamma^2} \,,
    \label{muspert}
\end{equation}
where
\begin{equation}
    c_1 = \frac{\mu _I^2+2 (\cos \theta+1) m_K^2-m_{\pi }^2 (\cos \theta+1)}{\mu _I \left(c_2+\mu _I\right)+2 m_K^2-m_{\pi }^2}, \quad c_2 = \sqrt{\mu _I^2+4 m_K^2-2 m_{\pi }^2} \,.
\end{equation}
Note that for realistic values of the meson masses the $\cO(\gamma)$ correction to $\mu_{s,cr}$ is quite large, despite the tiny value of $\gamma$. This fact is illustrated in Fig. \ref{cumpa} where we compare the exact value (obtained numerically) of $\mu_{s,cr}^{(PK)}$ as a function of $\theta$ to the leading order of its small $\gamma$ expansion for $m_\pi = 140 $ MeV and $m_K = 495 $ MeV. Note that, unlike the degenerate quarks case of Sec. \ref{masseuguali}, $\mu_{s,cr}^{(PK)}$ depends nontrivially from $\theta$. As a consequence, the Pion to Kaon transition may be induced by a change of $\theta$. This is exemplified in Fig. \ref{thetachange} where we show the jump of the $\expval{\bar u u}$ condensate across the transition.

\begin{figure}[t!]
\centering
\includegraphics[width=0.85\textwidth]{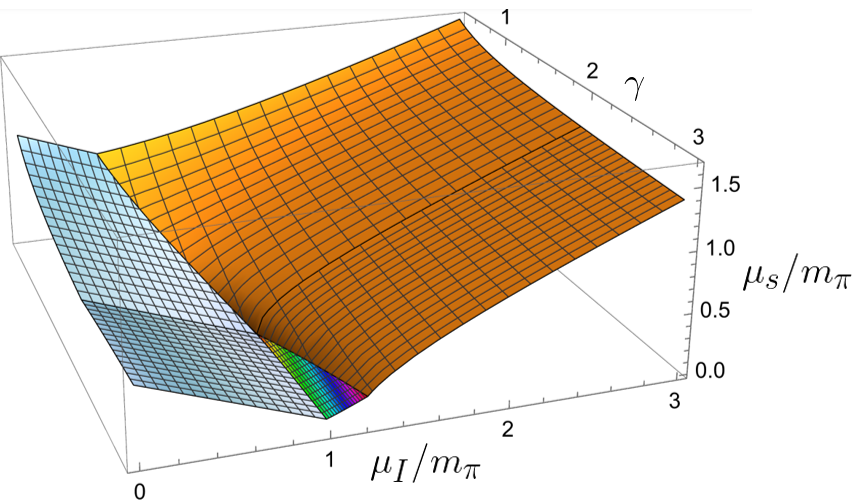} 
	% \caption{The Kaon phase is realized for values on $\mu_s$ above the depicted surfaces for $\theta=\pi$. Across the orange and rainbow surfaces the transition is first order and occurs between the Pion and the Kaon phase. The light blue surface separates Normal and Kaon phases and represents a second order phase transition. The rainbow surface represents the region $\gamma \ge 2$, $\mu_I \le \mu_I^*$ where the Pion phase breaks parity. The intersection between rainbow and light blue surfaces and between orange and light blue surfaces defines the quadruple line.} 
    \caption{The Kaon phase is realized for values on $\mu_s$ above the depicted surfaces for $\theta=\pi$. Across the orange surface, the transition is first order and occurs between the Kaon phase and the superfluid phase characterized by $\expval{\bar u d} \neq 0$ and vanishing pion condensate. The light blue surface separates the Normal and Kaon phases, representing a second-order phase transition. The rainbow surface corresponds to the first order transition separating the Pion and Kaon phases. The intersection between rainbow and light blue surfaces and orange and light blue surfaces defines the quadruple line. Finally, the intersection between orange and rainbow surfaces corresponds to a tricritical line.} 
	\label{nus}
\end{figure}

At $\theta=\pi$ the Kaon phase is realized when the following conditions hold
\begin{itemize}
    \item[]  $\bullet \ \gamma < 2$ \be
    \mu_s >  \sqrt{m_K^2-\frac{m_\pi^2}{2}}-\frac{\mu_I}{2} \,, \quad \text{and} \quad \mu_s > \sqrt{\frac{\mu_I^2}{4}+m_K^2-m_\pi^2} \,,
    \ee
      \item[] $\bullet \ \gamma \ge 2$ and $\mu_I \le \mu_I^*$ \be
    \mu_s >  \sqrt{m_\pi^2- m_K^2}-\frac{\mu_I}{2} \,, \quad \text{and} \quad \mu_s > \frac{\sqrt{\left(\mu_I^2+m_\pi^2\right)^2-4 \mu_I^2 m_K^2}-m_\pi^2}{2 \mu_I} \,,
    \ee
     \item[] $\bullet \ \gamma \ge 2$ and $\mu_I \ge \mu_I^*$ \be
    \mu_s >  \sqrt{m_\pi^2- m_K^2}-\frac{\mu_I}{2} \,, \quad \text{and} \quad \mu_s > \sqrt{\frac{\mu_I^2}{4}+m_K^2-m_\pi^2} \,.
    \ee
\end{itemize}
We illustrate this behavior in Fig. \ref{nus}. The above should be compared with the simple solution at $\theta=0$ \cite{Kogut:2001id}
\be
 \mu_s >  m_K-\frac{\mu_I}{2} \,, \quad \text{and} \quad  \mu_s >  \frac{\sqrt{\left(\mu_I^2-m_\pi^2\right)^2+4 \mu_I^2 m_K^2}-m_\pi^2}{2 \mu_I} \,.
\ee
Finally, the tricritical line where the three phases meet corresponds to
\begin{align}
    \mu_I &=  m_\pi \sqrt{\cos \left(\frac{\theta -\alpha_3^{(\pm)}}{2}\right)} \,, \nonumber \\
    \mu_s = & \sqrt{m_K^2 \cos \alpha_3^{(\pm)}+\frac{1}{2} m_\pi^2 \left(\cos \left(\frac{\theta -\alpha_3^{(\pm)}}{2}\right)-\cos \alpha_3^{(\pm)}\right)}-\frac{1}{2} m_\pi \sqrt{\cos \left(\frac{\theta -\alpha_3^{(\pm)}}{2}\right)} \,,
\end{align}
which at $\theta = \pi$ becomes
\begin{align}
    \mu_I &= \frac{m_\pi^2}{\sqrt{4 m_K^2-2 m_\pi^2}} \,, \quad \mu_s = \sqrt{\frac{2}{2 m_K^2-m_\pi^2}} \left(m_K^2-\frac{3 m_\pi^2}{4}\right) \,, \ \ \ \text{for} \quad \gamma < 2 \,,  \nonumber \\
     \mu_I &=m_\pi\,, \qquad \qquad \qquad  \mu_s = \sqrt{m_\pi^2-m_K^2}-\frac{m_\pi}{2}\,, \qquad   \qquad  \quad  \ \ \text{for} \quad \gamma \ge  2  \,.
\end{align}

\section{Condensates and charge densities} \label{sezioconde}
Our goal in this section is to improve the description of the phase diagram of the theory by computing the various condensates and charge densities. To this aim, we adopt the usual procedure of introducing sources into the chiral Lagrangian and taking derivatives with respect to them. Moreover, we found it instructive to present results for the most general case where all three quark masses are different. 
Without losing generality, we set the angles $\xi$ and $\rho$ appearing in the vacuum ansatz and representing the leftover $U(1)_I$ and $U(1)_s$ symmetries as $\xi = \frac{\pi}{2}$ and $\rho = \frac{\pi}{2}-\frac{\alpha_1-\alpha_3}{2}$. This choice aligns the vacuum along $\expval{\bar{u} d}$ and $\expval{\bar{u} s}$ rather than $\expval{\bar{d} u}$ and $\expval{\bar{s} u}$ (and the same for the corresponding pseudoscalar condensates).
Our results for the quark condensates are listed in Table \ref{tabellacond}. To avoid overburdening the presentation, we specify that the $\cos\varphi$ appearing in Table \ref{tabellacond} refers to Eq. \eqref{cosphipionico} and Eq. \eqref{cosphikaonico} for the Pion and Kaon phases, respectively.  

\begin{longtable}[c]{cccc}
\hline
\textbf{Condensates}                                                       & \textbf{Normal phase}                 & \textbf{Pion phase}                                                             & \textbf{Kaon phase}                                                              \\ \hline
\endfirsthead
\endhead
\multicolumn{1}{|c|}{\cellcolor[HTML]{FFFC9E}$\expval{\bar{u} u}$}         & \multicolumn{1}{c|}{$F_\pi^2 G \cos\alpha_1$} & \multicolumn{1}{c|}{$ F_\pi^2  G \cos\alpha_1\cos\varphi$}                                 & \multicolumn{1}{c|}{$ F_\pi^2 G \cos\alpha_1\cos\varphi$}                                  \\ \hline
\multicolumn{1}{|c|}{\cellcolor[HTML]{FFFC9E}$\expval{\bar{d} d}$}         & \multicolumn{1}{c|}{$ F_\pi^2 G \cos\alpha_2$} & \multicolumn{1}{c|}{$F_\pi^2  G \cos\alpha_2\cos\varphi$}                                 & \multicolumn{1}{c|}{$F_\pi^2  G \cos\alpha_2$}                                          \\ \hline
\multicolumn{1}{|c|}{\cellcolor[HTML]{FFFC9E}$\expval{\bar{s} s}$}         & \multicolumn{1}{c|}{$F_\pi^2  G \cos\alpha_3$} & \multicolumn{1}{c|}{$F_\pi^2  G \cos\alpha_3$}                                           & \multicolumn{1}{c|}{$F_\pi^2  G \cos\alpha_3\cos\varphi$}                                  \\ \hline
\multicolumn{1}{|c|}{\cellcolor[HTML]{FFFC9E}$\expval{\bar{u} d}$}         & \multicolumn{1}{c|}{$0$}              & \multicolumn{1}{c|}{$2 F_\pi^2  G \sin\left(\frac{\alpha_1+\alpha_2}{2}\right)\sin\varphi$} & \multicolumn{1}{c|}{$0$}                                                         \\ \hline
\multicolumn{1}{|c|}{\cellcolor[HTML]{FFFC9E}$\expval{\bar{u} s}$}         & \multicolumn{1}{c|}{$0$}              & \multicolumn{1}{c|}{$0$}                                                        & \multicolumn{1}{c|}{$-2 F_\pi^2  G \sin\left(\frac{\alpha_1+\alpha_3}{2}\right)\sin\varphi$} \\ \hline
% \multicolumn{1}{|c|}{\cellcolor[HTML]{FFFC9E}$\expval{\bar{d} s}$}         & \multicolumn{1}{c|}{$0$}              & \multicolumn{1}{c|}{$0$}                                                        & \multicolumn{1}{c|}{$0$}                                                         \\ \hline
\multicolumn{1}{|c|}{\cellcolor[HTML]{8ecae6}$\expval{\bar{u}\gamma_5 u}$} & \multicolumn{1}{c|}{$-F_\pi^2  G\sin\alpha_1$} & \multicolumn{1}{c|}{$-F_\pi^2 G\sin\alpha_1\cos\varphi$}                                   & \multicolumn{1}{c|}{$-F_\pi^2 G\sin\alpha_1\cos\varphi$}                                    \\ \hline
\multicolumn{1}{|c|}{\cellcolor[HTML]{8ecae6}$\expval{\bar{d}\gamma_5 d}$} & \multicolumn{1}{c|}{$-F_\pi^2  G\sin\alpha_2$} & \multicolumn{1}{c|}{$-F_\pi^2 G\sin\alpha_2\cos\varphi$}                                   & \multicolumn{1}{c|}{$-F_\pi^2 G \sin\alpha_2$}                                           \\ \hline
\multicolumn{1}{|c|}{\cellcolor[HTML]{8ecae6}$\expval{\bar{s}\gamma_5 s}$} & \multicolumn{1}{c|}{$-F_\pi^2 G\sin\alpha_3$} & \multicolumn{1}{c|}{$-F_\pi^2 G\sin\alpha_3$}                                           & \multicolumn{1}{c|}{$-F_\pi^2 G \sin\alpha_3\cos\varphi$}                                   \\ \hline
\multicolumn{1}{|c|}{\cellcolor[HTML]{8ecae6}$\expval{\bar{u}\gamma_5 d}$} & \multicolumn{1}{c|}{$0$}              & \multicolumn{1}{c|}{$2F_\pi^2 G\cos\left(\frac{\alpha_1+\alpha_2}{2}\right)\sin\varphi$}   & \multicolumn{1}{c|}{$0$}                                                         \\ \hline
\multicolumn{1}{|c|}{\cellcolor[HTML]{8ecae6}$\expval{\bar{u}\gamma_5 s}$} & \multicolumn{1}{c|}{$0$}              & \multicolumn{1}{c|}{$0$}                                                        & \multicolumn{1}{c|}{$-2F_\pi^2 G\cos\left(\frac{\alpha_1+\alpha_3}{2}\right)\sin\varphi$}   \\ \hline
% \multicolumn{1}{|c|}{\cellcolor[HTML]{DAE8FC}$\expval{\bar{d}\gamma_5 s}$} & \multicolumn{1}{c|}{$0$}              & \multicolumn{1}{c|}{$0$}                                                        & \multicolumn{1}{c|}{$0$}                                                         \\ \hline
\caption{Table of scalar (in yellow) and pseudo-scalar (in blue) quark condensates in the Normal, Pion, and Kaon phases.}
\label{tabellacond}\\
\end{longtable}
As expected the pion and kaon condensates appear only in the respective superfluid phases. Moreover, for nonzero values of the $\theta$-angle the CP-odd $\braket{\bar u \gamma_5 u}$, $\braket{\bar d \gamma_5 d}$ and CP-even $\braket{\bar u d}$, $\braket{\bar u s}$ condensates form \footnote{We recall that for $\theta=0$ one has $\alpha_1=\alpha_2=\alpha_3 = 0$.}. The condensates exhibit discontinuities across the first order transitions such as the one separating Pion and Kaon phases and the parity-restoring transition at $\theta = \pi$ and $\mu_I = \mu_I^*$. 

\begin{figure}[t!]
\centering
\includegraphics[width=0.45\textwidth]{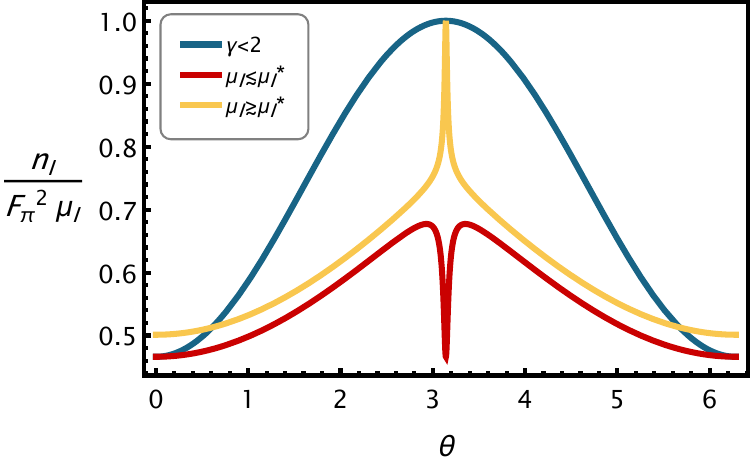} \quad  \includegraphics[width=0.47\textwidth]{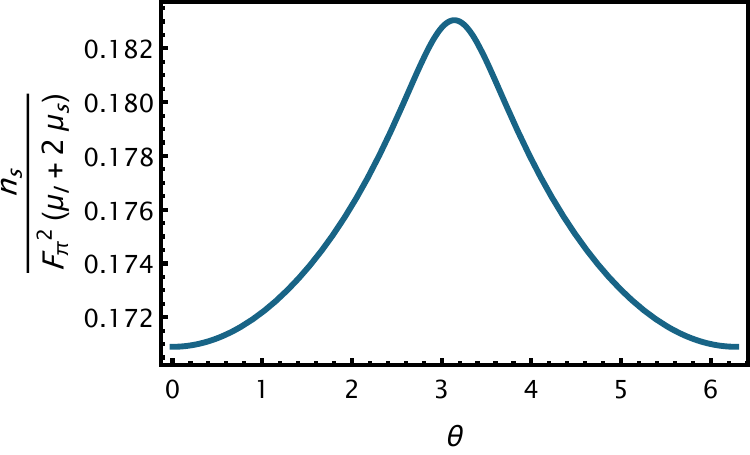} 
	\caption{{\it Left panel}: the isospin charge density $n_I$ as a function of $\theta$ in the Pion phase with degenerate masses for the up and down quarks. The blue line corresponds to the physical value of the meson masses $m_\pi=140 $ MeV and $m_K = 495 $ MeV for which $\gamma \sim 0.04$. The yellow and red lines have been obtained for $\gamma =2.8$ and $\mu_I$ just above and below $\mu_I^* \sim 165 $ MeV, respectively, to show the discontinuity across the parity restoring transition at $\theta=\pi$. {\it Right panel}: the strangeness charge density $n_s$ as a function of the $\theta$ in the Kaon phase for $m_\pi=140 $ MeV, $m_K = 495 $ MeV, $\mu_I= 280 $ MeV, and $\mu_s=520 $ MeV.} 
	\label{niens}
\end{figure}

The isospin $ n_I \equiv \pdv{\mathcal{L}}{\mu_I}$ and strangeness $ n_s \equiv  \pdv{\mathcal{L}}{\mu_s}$ number densities read
\begin{align}
    n_I &= F^2_\pi \mu_I \left(1-\frac{G^2\left(m_u\cos\alpha_1+m_d\cos\alpha_2\right)^2}{\mu^4_I}\right) \,, \qquad n_s = 0 \,,&\text{Pion phase} \,, \\
    n_I &= \frac{1}{4} F^2_\pi (\mu_I+2\mu_s) \left(1-\frac{16 G^2\left(m_u\cos\alpha_1+m_s\cos\alpha_3\right)^2}{(\mu_I+2\mu_s)^4}\right) \,, \quad n_s = 2 n_I \,, &\text{Kaon phase} \,,
\end{align}
and their $\theta$-dependence is illustrated in Fig. \ref{niens}.
Additionally, the Pion and Kaon phases feature nonzero values of certain $U(1)$ charge densities. Their expressions and values are listed in Table \ref{tabellavettass}. 

\begin{longtable}[c]{ccc}
\hline
\textbf{$U(1)$ charge densities}                                                                 & \textbf{Pion phase}            & \textbf{Kaon phase}      \\ \hline
\endfirsthead
\endhead
\multicolumn{1}{|c|}{\cellcolor[HTML]{e56b6f}$\expval{\bar{u}\gamma_0 d}$}           & \multicolumn{1}{c|}{$-F^2_\pi \mu_I \sin\left(\frac{\alpha_1-\alpha_2}{2}\right)\sin(2\varphi)$} & \multicolumn{1}{c|}{$0$} \\ \hline
\multicolumn{1}{|c|}{\cellcolor[HTML]{e56b6f}$\expval{\bar{u}\gamma_0 s}$}  &
  \multicolumn{1}{c|}{$0$} &
  \multicolumn{1}{c|}{$-\frac{1}{2}F^2_\pi(\mu_I+2\mu_s)\sin\left(\frac{\alpha_1-\alpha_3}{2}\right)\sin(2\varphi)$} \\ \hline
% \multicolumn{1}{|c|}{\cellcolor[HTML]{FFCCC9}$\expval{\bar{d}\gamma_0 s}$}          & \multicolumn{1}{c|}{$0$}       & \multicolumn{1}{c|}{$0$} \\ \hline
\multicolumn{1}{|c|}{\cellcolor[HTML]{b5e48c}$\expval{\bar{u}\gamma_0 \gamma_5 d}$}  &
  \multicolumn{1}{c|}{$F^2_\pi\mu_I\cos\left(\frac{\alpha_1-\alpha_2}{2}\right)\sin(2\varphi)$} &
  \multicolumn{1}{c|}{$0$} \\ \hline
\multicolumn{1}{|c|}{\cellcolor[HTML]{b5e48c}$\expval{\bar{u}\gamma_0 \gamma_5 s}$}  & \multicolumn{1}{c|}{$0$}       & \multicolumn{1}{c|}{$-\frac{1}{2}F^2_\pi(\mu_I+2\mu_s)\cos\left(\frac{\alpha_1-\alpha_3}{2}\right)\sin(2\varphi)$} \\ \hline
% \multicolumn{1}{|c|}{\cellcolor[HTML]{DDDDDD}$\expval{\bar{d}\gamma_0 \gamma_5 s}$}  & \multicolumn{1}{c|}{$0$}       & \multicolumn{1}{c|}{$0$} \\ \hline
\caption{$U(1)$ charge densities in the Pion, and Kaon phases.}
\label{tabellavettass}\\
\end{longtable}
Interestingly, the axial charge densities $\expval{\bar{u}\gamma_0 \gamma_5 d}$ and $\expval{\bar{u}\gamma_0 \gamma_5 s}$ are nonzero also for $\theta=0$ and act as an additional order parameter for the Pion and Kaon phases, respectively. This was already noted for the Pion phase with $N_f=2$ in \cite{Metlitski:2005di}.

We conclude the section by investigating the impact of the $\theta$-angle on the gluon condensate $\expval{G^2}$. The latter can be determined by starting from the expression for the trace anomaly 
\begin{equation}\label{trace}
    T^\mu_\mu =\frac{\beta(\alpha_s)}{4 \alpha_s} G_{\mu\nu}^a G^{\mu\nu a}+\bar{\psi}_I M \psi_I \,.
\end{equation}
Here $\beta(\alpha_s)$ is the beta function of the strong coupling $\alpha_s$ and $ T^\mu_\mu$ denotes the trace of the energy-momentum tensor of the theory which reads
\begin{equation}
    T^\mu_\mu = 3 p_i- E_i \,, \qquad i=N,P,K \,,
\end{equation}
where the pressure $p_i$ is related to the grand canonical potential $\Omega_i$ of the phase $i$ as $p_i=-\Omega_i$. Explicitly, we have
\begin{align}
    \Omega_N = E_N  \,, \qquad
    \Omega_P = E_P + \mu_I n_I  \,, \qquad
    \Omega_K = E_K + \mu_I n_I  + \mu_s n_s \ .
\end{align}
Accordingly, using the results obtained in this section and Eq. \eqref{trace}, we arrive at the expressions for the gluon condensate in the various phases
\begin{align}
\expval{\frac{\beta(\alpha_s)}{4 \alpha_s} G_{\mu\nu}^a G^{\mu\nu a}}_N &= 3 G F^2_\pi \left(m_u\cos\alpha_1+m_d\cos\alpha_2+m_s\cos\alpha_3\right) \,, \\
\expval{\frac{\beta(\alpha_s)}{4 \alpha_s} G_{\mu\nu}^a G^{\mu\nu a}}_P &= F^2_\pi\mu_I^2\left(1+2\cos^2\varphi\right)+3F^2_\pi G m_s\cos\alpha_3 \,, \\
    \expval{\frac{\beta(\alpha_s)}{4 \alpha_s} G_{\mu\nu}^a G^{\mu\nu a}}_K &= \frac{F^2_\pi}{4}(\mu_I+2\mu_s)^2\left(1+2\cos^2\varphi\right)+3F^2_\pi G m_d \cos\alpha_2\cos\varphi \,.
\end{align}

\section{Spectrum} \label{spectrum}

Having discussed the impact of the $\theta$ angle on the phase diagram and the condensates, the time is ripe to determine the mass spectrum of the theory. The symmetry breaking patterns for the Pion and Kaon phases are
\vskip 1em
\centerline{ \bf Pion Phase}

\begin{align}
   SU(3)_L \times SU(3)_R \times U(1)_V  \overset{\mu_I,\mu_s}{\longrightarrow} U(1)_I \times U(1)_S \times U(1)_{V=B} \overset{\Sigma_I}{\rightsquigarrow}  U(1)_S \times U(1)_{V=B} \,.    
\end{align}

\centerline{ \bf Kaon Phase}

\begin{align}
 SU(3)_L \times SU(3)_R \times U(1)_V  \overset{\mu_I,\mu_s}{\longrightarrow} U(1)_I \times U(1)_S \times U(1)_{V=B} \overset{\Sigma_S}{\rightsquigarrow} U(1)_I \times U(1)_{V=B} \,.    
\end{align}

We, therefore, expect one single gapless Goldstone mode in each superfluid phase. The calculation of the mass spectrum for $\theta=0$ first appeared in \cite{Kogut:2001id}, has later been corrected in \cite{Mammarella:2015pxa, Adhikari:2019mlf}, and will be reviewed below. For the sake of simplicity, we assume $m_u = m_d = m$. 

\subsection{Spectrum for $\theta = 0$} \label{notheta}

To obtain the mass spectrum at zero $\theta$-angle, we set to zero the last term in \eqref{lagtheta} and expand the Lagrangian to the quadratic order in the fluctuations which we parameterize in terms of real fields as follows
\begin{equation}
    \Sigma=e^{i\frac{\Phi}{\sqrt{2}F_\pi}} \Sigma_c e^{i\frac{\Phi}{\sqrt{2}F_\pi}} \,, \quad\text{where}\quad \Phi = \begin{pmatrix} \label{fluc}
        \frac{\pi_3}{\sqrt{2}}+\frac{\pi_8}{\sqrt{6}} & \frac{\pi_1-i \pi_2}{\sqrt{2}} & \frac{\pi_4-i \pi_5}{\sqrt{2}}\\
        \frac{\pi_1+i \pi_2}{\sqrt{2}} & -\frac{\pi_3}{\sqrt{2}}+\frac{\pi_8}{\sqrt{6}} & \frac{\pi_6-i \pi_7}{\sqrt{2}}\\
        \frac{\pi_4+i \pi_5}{\sqrt{2}} & \frac{\pi_6+i \pi_7}{\sqrt{2}} & -\sqrt{\frac{2}{3}}\pi_8
    \end{pmatrix}\ ,
\end{equation}
with $\Sigma_c$ in Eq. \eqref{SadP}. We assume the $a \gg m_\pi^2, m_K^2$ limit and, therefore, exclude the singlet mode $S$. Without losing generality, we set $\xi=\rho = \pi/2$. 

In the Normal phase, the quadratic Lagrangian in momentum space can be written as
\begin{equation}
    \mathcal{L} = \Pi D^{-1}\ \Pi^T \,,
\end{equation}
where $\Pi=\{\pi_1,\pi_2,\pi_3,\pi_4,\pi_5,\pi_6,\pi_7,\pi_8\}$ and the inverse propagator $D^{-1} $ takes a block diagonal form. As usual, the meson masses are obtained by solving  $\det{D^{-1}}=0$ for vanishing spatial momentum $k$  and read 
\begin{equation}\label{spettronormale}
\begin{aligned}
    m_{\pi_0}&=m_\pi \,,\\
    m_{\pi_+}&=m_\pi-\mu_I \,,\\
    m_{\pi_-}&=m_\pi+\mu_I \,,\\
    m_{K_+}&=m_K-\frac{1}{2}\mu_I-\mu_s\,,\\
    m_{K_-}&=m_K+\frac{1}{2}\mu_I+\mu_s\,,\\
    m_{K_0}&=m_K+\frac{1}{2}\mu_I-\mu_s\,,\\
    m_{\bar{K}_0}&=m_K-\frac{1}{2}\mu_I+\mu_s\,,\\
    m_\eta&=\sqrt{\frac{4m^2_K-m^2_\pi}{3}} \,.
\end{aligned}
\end{equation}

In the Pion phase, the kinetic term in the Lagrangian can be brought into the canonical form by performing the following field redefinitions \cite{Mammarella:2015pxa}
\begin{align}
    \tilde{\pi}_{1,3}=\pi_{1,3}\cos\varphi\,, \qquad
\tilde{\pi}_{4,5,6,7}=\pi_{4,5,6,7}\cos\left(\frac{\varphi}{2}\right)\,,
\end{align}
with $\cos \varphi = \frac{m_\pi^2}{\mu_I^2}$. Hence, by setting $\beta = 0$ in Eq. \eqref{SadP}, one obtains
\begin{equation}
    \begin{aligned}
        m_{\tilde{\pi}_0}&=\mu_I \,,\\
        m_{\tilde{\pi}_+}&=0 \,,\\
        m_{\tilde{\pi}_-}&= \frac{\sqrt{3m_\pi^4+\mu_I^4}}{\mu_I} \,,\\
        m_{\tilde{K}_+}& =\frac{\sqrt{-2 \mu _I^2 \left(m_{\pi }^2-2 m_K^2\right)+\mu _I^4+m_{\pi }^4}-2 \mu _I \mu _s-m_{\pi }^2}{2 \mu _I} \,,\\
        m_{\tilde{K}_-}& =\frac{\sqrt{-2 \mu _I^2 \left(m_{\pi }^2-2 m_K^2\right)+\mu _I^4+m_{\pi }^4}+2 \mu _I \mu _s+m_{\pi }^2}{2 \mu _I} \,,\\
        m_{\tilde{K}_0}& =\frac{\sqrt{-2 \mu _I^2 \left(m_{\pi }^2-2 m_K^2\right)+\mu _I^4+m_{\pi }^4}-2 \mu _I \mu _s+m_{\pi }^2}{2 \mu _I} \,,\\
         m_{\tilde{\bar{K}}_0}& =\frac{\sqrt{-2 \mu _I^2 \left(m_{\pi }^2-2 m_K^2\right)+\mu _I^4+m_{\pi }^4}+2 \mu _I \mu _s-m_{\pi }^2}{2 \mu _I} \,,\\
       m_{\tilde{\eta}} & = \frac{\sqrt{m_\pi^4-2(m_\pi^2-2 m_K^2)\mu_I^2}}{\sqrt{3}\mu_I}\ .
    \end{aligned}
\end{equation}
Here $\tilde{\pi}_+$ is the Goldstone boson stemming from the spontaneous breaking of $U(1)_I$. 

Finally, in the Kaon phase, the canonical form of the quadratic Lagrangian is obtained by redefining the fields as follows \cite{Mammarella:2015pxa}
\begin{align}
    \hat{\pi}_{1,2,6,7}=\pi_{1,2,6,7}\cos\left(\frac{\varphi}{2}\right) \,, \quad
\hat{\pi}_{4}=\pi_{4}\cos\varphi\ \,, \quad
\begin{pmatrix}
    \hat{\pi}_3\\
    \hat{\pi}_8
\end{pmatrix}  = \begin{pmatrix}
    -\frac{\sqrt{3}}{2} & \frac{1}{2}\\
    \frac{1}{2}\cos\varphi & \frac{\sqrt{3}}{2}\cos\varphi
\end{pmatrix}\begin{pmatrix}
    \pi_3\\
    \pi_8
\end{pmatrix}\,,
\end{align}
with $\cos \varphi =\frac{4m_K^2}{\left(\mu_I + 2\mu_s \right)^2}$. Therefore, by fixing $\beta=\pi/2$ we arrive at 
\allowdisplaybreaks
\begin{align} \label{ksp0}
        m^2_{\hat{\pi}_0}&=\frac{1}{6} \left(\frac{4 m_K^4}{\left(\mu _I+2 \mu _s\right)^2}-M_{38}+\frac{3 \mu _I^2}{4}+3 \mu _I \mu _s+2 m_{\pi }^2+3 \mu _s^2\right) \,,\\
        m^2_{\hat{\pi}_+}&=\frac{8 m_K^2 \left(\mu _I^2-4 \mu _s^2\right)- M_{12}+\left(\mu _I+2 \mu _s\right)^2 \left(5 \mu _I^2-4 \mu _I \mu _s+8 m_{\pi }^2+4 \mu _s^2\right)+16 m_K^4}{8 \left(\mu _I+2 \mu _s\right)^2}  \,,\\
        m^2_{\hat{\pi}_-}&= \frac{8 m_K^2 \left(\mu _I^2-4 \mu _s^2\right)+M_{12}+\left(\mu _I+2 \mu _s\right)^2 \left(5 \mu _I^2-4 \mu _I \mu _s+8 m_{\pi }^2+4 \mu _s^2\right)+16 m_K^4}{8 \left(\mu _I+2 \mu _s\right)^2}  \,,\\
        m^2_{\hat{K}_+}& =0  \,,\\
        m^2_{\hat{K}_-}& =\frac{\left(\mu _I+2 \mu _s\right)^4+48 m_K^4}{4 \left(\mu _I+2 \mu _s\right)^2}  \,,\\
        m^2_{\hat{K}_0}& =\mu_I^2  \,,\\
        m^2_{\hat{\bar{K}}_0} &=\frac{\left(\mu _I^2-4 m_K^2-4 \mu _s^2\right)^2}{4 \left(\mu _I+2 \mu _s\right)^2}  \,,\\
       m^2_{\hat{\eta}} & = \frac{1}{6} \left(\frac{4 m_K^4}{\left(\mu _I+2 \mu _s\right)^2}+M_{38}+\frac{3 \mu _I^2}{4}+3 \mu _I \mu _s+2 m_{\pi }^2+3 \mu _s^2\right)  \,,
\end{align}
where  
{\small
\begin{align}
 M_{12}&=\sqrt{\left[4m^2_K+(3\mu_I-2\mu_s)(\mu_I+2\mu_s) \right]^2\left(16 m_{\pi }^2 \left(\mu _I+2 \mu _s\right)^2 -8 m_K^2 \left(\mu _I+2 \mu _s\right)^2+\left(\mu _I+2 \mu _s\right)^4+16 m_K^4\right)}  \,,\\
    M_{38}&=\sqrt{\left(\frac{3 \left(\mu _I+2 \mu _s\right)^4+16 m_K^4}{4 \left(\mu _I+2 \mu _s\right)^2}+2 m_{\pi }^2\right)^2-6 m_{\pi }^2 \left(\left(\mu _I+2 \mu _s\right)^2+4 m_K^2-2 m_{\pi }^2\right)}
   \ .
\end{align}}
The gapless mode $ m^2_{\hat{K}_+}$ is the Goldstone boson arising from the SSB of $U(1)_S$.

\subsection{Spectrum for $\theta \neq 0$}

We now derive the spectrum of the theory at nonzero $\theta$-angle. The parameterization of the fluctuations is again given by Eq. \eqref{fluc} upon replacing the $\theta= 0$ vacuum $\Sigma_c$ with $\Sigma_0$ in Eq. \eqref{eq:ansatsvacuumcomplete}. Since for $\theta \neq 0$ the meson masses are given by rather cumbersome expressions, in what follows, we will instead report the blocks of the inverse propagator from which it is straightforward to extract the spectrum.

\begin{figure}[t!]
    \centering
    \subfloat[ Normal phase for $\mu_I=20$  and $\mu_s=50$. Note that at $\theta=\pi$ the $\pi_+$ mode becomes almost gapless. This can be traced back to the tiny value of $\mu_{I,cr}(\theta=\pi)$ for the considered values of $m_\pi$ and $m_K$, as can be seen from Eq. \eqref{tino}.]{\includegraphics[scale=0.615]{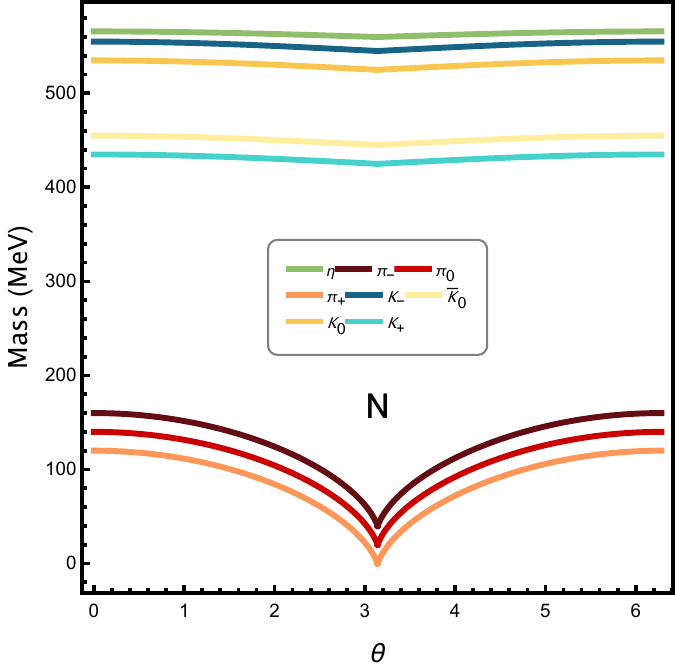} }%
    \vspace{1em}
    \subfloat[\centering  Pion phase for $\mu_I=150$ and $\mu_s=50$ MeV. ]{\includegraphics[scale=0.55]{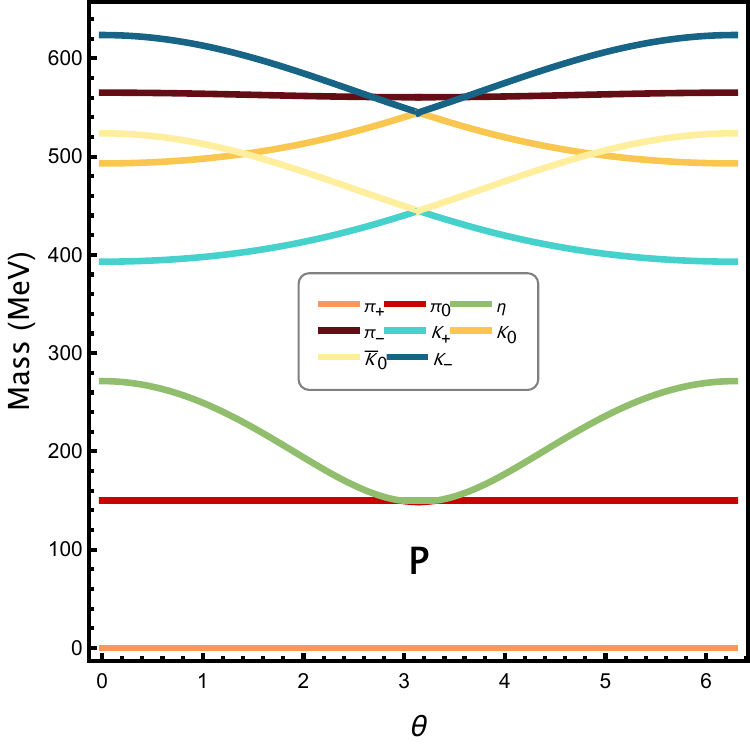} }%
     \vspace{1em}
    \subfloat[\centering Kaon phase for $\mu_I=10$ and $\mu_s=520$.]{\includegraphics[scale=0.45]{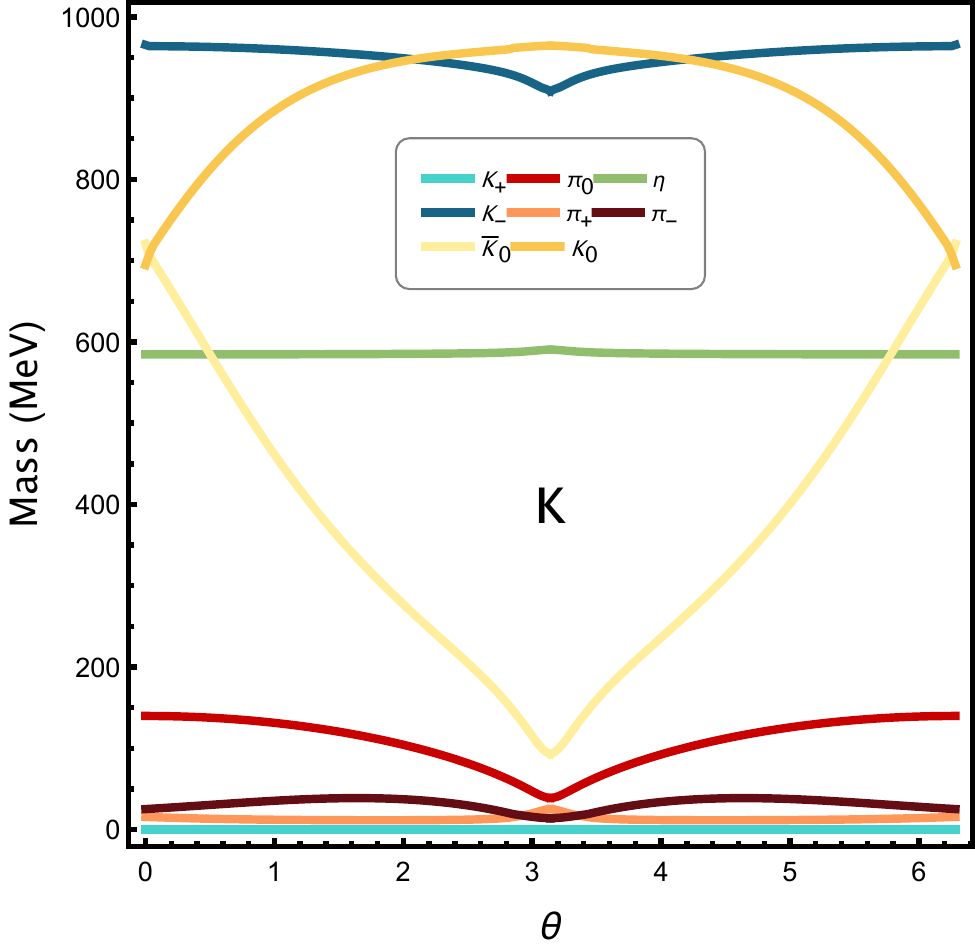}}
     \caption{Masses of the meson octet as a function of $\theta$ for $m_\pi=140$ MeV, $m_K=495$ MeV.}%
    \label{fig:fasiQCD2}%
\end{figure}

In the Normal phase, the inverse propagator is block diagonal and can be written as

\begin{equation} \label{blocco1}
D^{-1}=\begin{pNiceArray}{ccc|ccccc}
  \Block{2-2}<\large>{D_{12}} & & 0 & \Block{3-5}<\LARGE>{\mathbf{0}}  &  &  &  &   \\
 && \Vdots &   & & &  &  \\
  0 & \Cdots & D_{33} &  &  &   &  &  \\
  \hline
  \Block{5-3}<\LARGE>{\mathbf{0}} && &  \Block{2-2}<\large>{D_{45}} && \Block{2-2}<\Large>{\mathbf{0}} &&0 \\
   & &   &  &  &  &  &\Vdots \\
   && &  \Block{2-2}<\Large>{\mathbf{0}} && \Block{2-2}<\large>{D_{67}} && \Vdots\\
   & &   &  &  &  &  & 0\\
    &&  &  0&  \Cdots& \Cdots & 0 &  D_{88}
\end{pNiceArray} \,,
\end{equation}
where the submatrices are
\begin{align} \label{subnormal}
D_{12} &= \begin{pmatrix}
   \omega^2-k^2+\mu_I^2-m_\pi^2\cos\alpha & -2 i \mu_I \omega \\
   2 i \mu_I \omega  &\omega^2-k^2+\mu_I^2-m_\pi^2\cos\alpha
\end{pmatrix} \,, \nonumber \\
D_{33} &= \omega^2-k^2-m_\pi^2\cos\alpha \,, \nonumber \\
D_{45} &= \begin{pmatrix}
  \omega^2-k^2 +\frac{1}{4}\left(\mu_I+2\mu_s\right)^2-\frac{m_K^2\left(\cos\alpha+\cos\alpha_3\right)}{1+\cos(\alpha-\alpha_3)} & -i (\mu_I+2\mu_s)\omega\\
    i (\mu_I+2\mu_s)\omega & \omega^2-k^2 +\frac{1}{4}\left(\mu_I+2\mu_s\right)^2-\frac{m_K^2\left(\cos\alpha+\cos\alpha_3\right)}{1+\cos(\alpha-\alpha_3)}
\end{pmatrix} \,,  \nonumber  \\
D_{67} &= \begin{pmatrix}
   \omega^2-k^2 +\frac{1}{4}\left(\mu_I-2\mu_s\right)^2-\frac{m_K^2\left(\cos\alpha+\cos\alpha_3\right)}{1+\cos(\alpha-\alpha_3)} & i(\mu_I-2\mu_s)\omega\\
    -i(\mu_I-2\mu_s)\omega &  \omega^2-k^2 +\frac{1}{4}\left(\mu_I-2\mu_s\right)^2-\frac{m_K^2\left(\cos\alpha+\cos\alpha_3\right)}{1+\cos(\alpha-\alpha_3)}
\end{pmatrix} \,, \nonumber  \\
    D_{88}&= \omega^2-k^2-\frac{1}{3}m_\pi^2\cos\alpha+\frac{2}{3}(m_\pi^2-2m_K^2)\cos\alpha_3 \,,
\end{align}
with $\alpha$ and $\alpha_3$ given in Eq. \eqref{baralfa} and Eq.\eqref{solnormal}, respectively. The indices that label the blocks refer to the corresponding field fluctuations $\pi_a$, $a=1, \dots, 8$. For $\gamma=2$ and $\theta=\pi$, the $\pi_8$ mode becomes gapless signaling that the Dashen phase transition becomes of the second order.  In Fig. \ref{fig:fasiQCD2}(a) we show the meson masses as a function of $\theta$ for the reference values $m_\pi =140$ MeV, $m_K =495$ MeV, $\mu_I =20$ MeV, and $\mu_s =50$ MeV.

In the Pion phase, the inverse propagator takes the form
\begin{equation}
D^{-1}=\begin{pNiceArray}{c @{\hspace{5mm}}c @{\hspace{5mm}}c @{\hspace{5mm}} |c @{\hspace{5mm}} c @{\hspace{5mm}}c @{\hspace{5mm}}c @{\hspace{5mm}}c @{\hspace{5mm}}}
  \Block{3-3}<\Large>{D_{128}} & & & \Block{3-5}<\LARGE>{\mathbf{0}}  &  &  &  &   \\
 &&  &   & & &  &  \\
   &  & &  &  &   &  &  \\
  \hline
  \Block{5-3}<\LARGE>{\mathbf{0}} && &  D_{33} & \Block{1-2}<\Large>{\mathbf{0}} &&\Block{1-2}<\Large>{\mathbf{0}} & \\
   && &  \Block{2-1}<\Large>{\mathbf{0}}  & \Block{4-4}<\LARGE>{D_{4567}} && &\\
   &&  &  &  &  &    &\\
    &&  & \Block{2-1}<\Large>{\mathbf{0}}  &  &  &    &\\
    & &   &  &  &  &  & \\
\end{pNiceArray} \,,
\end{equation}
where
\allowdisplaybreaks
{\begin{equation} \label{lostesso}\begin{aligned}
D_{128} & = \begin{pmatrix}
    \omega^2-k^2 & -2 i \mu_I \omega\cos\varphi & 0\\
    2 i \mu_I \omega\cos\varphi & \omega^2-k^2-m_\pi^2\cos\alpha\cos\varphi+\mu_I^2\cos(2\varphi) & -\frac{m_\pi^2\sin\alpha\sin\varphi}{\sqrt{3}}\\
    0 & -\frac{m_\pi^2\sin\alpha\sin\varphi}{\sqrt{3}} & D_{88}
\end{pmatrix} \,, \\
D_{33} &= \omega^2-k^2-\mu_I^2 \,, \\
D_{4567}&=\begin{pmatrix}
       D_{+} & 2i\mu_s\omega & -i \mu_I \omega\cos\varphi & - \mu_I \mu_s \cos\varphi\\
       -2i\mu_s\omega & D_{+} & - \mu_I \mu_s \cos\varphi & i \mu_I \omega\cos\varphi\\
       i \mu_I \omega\cos\varphi & - \mu_I \mu_s \cos\varphi & D_{-} & -2i\mu_s\omega\\
       - \mu_I \mu_s \cos\varphi & -i \mu_I \omega\cos\varphi & 2i\mu_s\omega & D_{-}
    \end{pmatrix} \,,
\end{aligned} \end{equation}}
with 
\begin{align}
    D_{88} & =\omega^2-k^2+\frac{2}{3}(m_\pi^2-2m_K^2)\cos\alpha_3-\frac{1}{3}m_\pi^2\cos\alpha\cos\varphi\,,\\
     D_{\pm} &= \omega^2-k^2+\frac{\abs{\cos(\alpha-\alpha_3)+\cos\varphi}\left(4\mu_s^2+\mu_I^2\cos\left(\frac{\alpha-\alpha_3\pm3\varphi}{2}\right)\sec\left(\frac{\alpha-\alpha_3\mp3\varphi}{2}\right)\right)}{4\sqrt{(1+\cos(\alpha-\alpha_3-\varphi))(1+\cos(\alpha-\alpha_3+\varphi))}} \nonumber \\ & -\frac{m_K^2(\cos\alpha+\cos(\alpha_3\mp\varphi))}{\cos(\alpha-\alpha_3)+\cos\varphi} \,,
\end{align}
and $\alpha$ and $\alpha_3$ respectively given in Eq. \eqref{baralfa} and Eq.\eqref{alpha3pion}. The $D_{128}$ block describes two massive modes and a massless Goldstone boson associated with the spontaneous breaking of $U(1)_I$. In Fig. \ref{fig:fasiQCD2}(b) we illustrate the $\theta$-dependence of the meson masses in the Pion phase for $m_\pi =140$ MeV, $m_K =495$ MeV, $\mu_I =150$ MeV, and $\mu_s =50$ MeV.

Finally, in the Kaon phase, the inverse propagator reads
\begin{equation}
D^{-1}=\begin{pNiceArray}{c @{\hspace{5mm}}c @{\hspace{5mm}}c @{\hspace{5mm}} c @{\hspace{5mm}}| c @{\hspace{5mm}}c @{\hspace{5mm}}c @{\hspace{5mm}}c @{\hspace{5mm}}}
  \Block{4-4}<\LARGE>{D_{1267}} & & & & \Block{4-4}<\LARGE>{\mathbf{0}}  &  &  &     \\
 &&  &   & & &  &  \\
   &  & &  &  &   &  &  \\
   &  & &  &  &   &  &  \\
  \hline
  \Block{4-4}<\LARGE>{\mathbf{0}} && &   & \Block{4-4}<\LARGE>{D_{3458}} & & &\\
   &&  &   & & &  &  \\
   &  & &  &  &   &  &  \\
   &  & &  &  &   &  &  \\
\end{pNiceArray} \,.
\end{equation}
The explicit expressions of the $4 \times 4$ blocks $D_{1267}$ and $D_{3458}$ are too involved to report here and can be provided upon request. We checked that the $D_{3458}$ block includes the expected Goldstone mode associated with the breaking of $U(1)_S$. Moreover, we illustrate the $\theta$-dependence of the meson masses in the Kaon phase in Fig. \ref{fig:fasiQCD2}(c) for $m_\pi =140$ MeV, $m_K =495$ MeV, $\mu_I =10$ MeV, and $\mu_s =520$ MeV.

% \FloatBarrier

% \FloatBarrier

\subsection{Spectrum at large $N_c$}

We conclude the section with a brief discussion of the spectrum in the large $N_c$ limit, where one has to include the additional Goldstone boson $S$ emerging from the spontaneous symmetry breaking of the $U(1)$ axial symmetry. In the Normal phase, the presence of the $S$ mode does not affect the $D_{12}, D_{33}, D_{45}$, and $D_{67}$  blocks in Eq. \eqref{subnormal}. On the other hand, $\pi_8$ mixes with the $S$ with the resulting masses being
 \begin{align}
 m^2_{8,9}= &\pm \frac{1}{2} \sqrt{9 a^2+\left(m_\pi^2 \cos \alpha - \left(2 m_K^2 - m_\pi^2\right)\cos \alpha_3 \right) \left(2 a + m_\pi^2 \cos \alpha - \left(2 m_K^2 - m_\pi^2\right)\cos \alpha_3\right)} \nonumber \\ &+\frac32 a + \frac{m_\pi^2}{2} \cos \alpha +\frac12 \left( 2m_K^2 - m_\pi^2 \right)\cos \alpha_3 \,.
 \end{align}
 
In the Pion phase, the $S$ mode mixes with both $\pi_2$ and $\pi_8$. Accordingly, the $D_{128}$ block of the inverse propagator in Eq. \eqref{lostesso} is replaced by a $4 \times 4$ submatrix $D_{1289}$ given by
\begin{equation}
D_{1289}=\begin{pNiceArray}{c @{\hspace{5mm}}c @{\hspace{5mm}} c @{\hspace{5mm}}c @{\hspace{5mm}}}
  \Block{2-2}<\large>{D_{12}} & & 0  & 0 \\
 &  &    -\frac{1}{\sqrt{3}} m_\pi^2\sin\alpha\sin\varphi & -2\sqrt{\frac23} m_\pi^2\sin\alpha\sin\varphi\\
  0 &  -\frac{1}{\sqrt{3}} m_\pi^2\sin\alpha\sin\varphi & \Block{2-2}<\large>{D_{89}} \\
  0 & -2\sqrt{\frac23} m_\pi^2\sin\alpha\sin\varphi &  &    \\
\end{pNiceArray} \,,
\end{equation}
where
{\footnotesize \begin{equation}
\begin{aligned}
D_{12} & = \begin{pmatrix}
  \omega^2-k^2 & -2 i \mu_I \omega\cos\varphi\\
    2 i \mu_I \omega\cos\varphi & \omega^2-k^2-m_\pi^2\cos\alpha\cos\varphi+\mu_I^2\cos(2\varphi)
\end{pmatrix} \,,\\
D_{89}&= \begin{pmatrix}
   \omega^2-k^2+\frac{2}{3}(m_\pi^2-2m_K^2)\cos\alpha_3-\frac{1}{3}m_\pi^2\cos\alpha\cos\varphi & \frac{\sqrt{2}}{3}[(2m_K^2-m_\pi^2)\cos\alpha_3-m_\pi^2\cos\alpha\cos\varphi]\\
    \frac{\sqrt{2}}{3}[(2m_K^2-m_\pi^2)\cos\alpha_3-m_\pi^2\cos\alpha\cos\varphi] & \omega^2-k^2+\frac{1}{3}(m_\pi^2-2m_K^2)\cos\alpha_3-\frac{2}{3}m_\pi^2\cos\alpha\cos\varphi-3 a
\end{pmatrix} \,.
\end{aligned} \end{equation}}

\section{Dashen's phenomenon for $N_f>3$} \label{generalNF}

We conclude our investigation by analyzing the phase diagram of the theory for $N_f>3$. As we shall see below, we will learn that the absence of Dashen's phenomenon is not a general property of the superfluid phases as the $N_f=3$ case may indicate. For the sake of simplicity, we restrict ourselves to the case of degenerate quark masses (i.e. $M= m \identity_{N_f}$) and a single chemical potential $\mu$ that we introduce by considering the following form of the $B$ matrix
\begin{equation}
\label{Massa}
 B = \frac{\mu}{2} \left(\begin{array}{ccc}
\identity_s & 0 & 0 \\ 0 & -\identity_s & 0 \\  0 & 0 & \mathbb{0}_{N_f-2s} 
\end{array}\right)  \,, \qquad s=1, \dots, N_f/2 \,.
\end{equation}
Note that for $s=1$, $\mu$ is the conventional isospin chemical potential.  The ground state takes the form 
\begin{align}
\label{eq:ansatsvacuum}
\Sigma_0 = W \Sigma_c \,, \qquad
 W  = \text{diag}\{e^{-i \alpha_1}\,, \dots \,,e^{-i \alpha_{N_f}} \} \,, \qquad \Sigma_c= \left(\begin{array}{ccc} \cos\varphi \ \mathbb{1}_s & \sin \varphi \ \mathbb{1}_s& \mathbb{0} \\ -\sin \varphi \ \mathbb{1}_s & \cos\varphi \ \mathbb{1}_s & \mathbb{0} \\
\mathbb{0} & \mathbb{0} & \mathbb{1}_{N_f-2s} \end{array} \right) \,,
% \qquad \text{and}\nonumber\\  \Sigma_I&= \left(\begin{array}{ccc} \mathbb{0}_s & \mathbbm{1}_s & 0 \\ \mathbbm{1}_s & \mathbb{0}_s & 0 \\
% 0 & 0 & \mathbb{0}_{N_f-2s} \end{array} \right)\ \cos\eta\  
% + i\left(\begin{array}{ccc} \mathbb{0}_s & -\mathbbm{1}_s & 0 \\ \mathbbm{1}_s & \mathbb{0}_s & 0 \\
% 0 & 0 & \mathbb{0}_{N_f-2s} \end{array} \right)\ \sin\eta \, . 
\end{align}
% It is useful to define the following quantities
% \begin{equation}
% X=\sum_{i=1}^{2s} \cos\alpha_i\ , \qquad Y=\sum_{i=2s+1}^{N_f} \cos\alpha_i \,,
% \end{equation}
and the Lagrangian evaluated on the ground state ansatz reads
\begin{equation}
    \mathcal{L}_0= \frac{F_\pi^2}{2} m_\pi^2 \left( \cos \varphi \sum_{i=1}^{2s} \cos\alpha_i+ \sum_{i=2s+1}^{N_f} \cos\alpha_i\right) +\frac{F_\pi^2}{2} s\mu ^2  \sin ^2 \varphi - \frac{a F_\pi^2}{4} \left( \sum^{N_f}_{i=1} \alpha_i - \theta \right)^2 \,,
\end{equation}
where $m_\pi$ denotes the tree-level meson mass. We again consider the $a \gg m_\pi^2$ limit corresponding to integrating out the $\eta^\prime$ state. Then the angle $\varphi$ and the Witten variables $\alpha_i$ are determined by the EOM
\begin{align} \label{thetavac}
    \sin \varphi  \left(\cos \varphi -\frac{m_\pi^2}{2 s \mu ^2}\sum_{i=1}^{2s} \cos\alpha_i\right)&=0 \,, \\ 
     m_\pi^2 \cos \varphi  \sin \alpha_i &= m_\pi^2 \sin \alpha_j \ , \,\,\, i=1,\dots, 2s\ , \, \, j=2s+1,\dots, N_f  \,,
\end{align}
supplemented by the constraint 
\be
\sum^{N_f}_{i=1} \alpha_i = \theta \,.
\ee
Equation \eqref{thetavac} has the solutions $\varphi= 0$ and $\cos\varphi =\frac{m_\pi^2}{2s\mu^2}\sum_{i=1}^{2s} \cos\alpha_i$ with the latter corresponding to a superfluid phase characterized by meson condensation. The energy of the theory in the two phases reads
\begin{align}
    E_N&=- \frac{1}{2} F_\pi^2  m_\pi^2 \sum_{i=1}^{N_f} \cos\alpha_i \,, \qquad \qquad \qquad \qquad \qquad \qquad \qquad  \quad \qquad\text{Normal phase} \,, \\
    E_S&=-\frac{1}{2} F_\pi^2\left( m_\pi^2 \sum_{i=2s+1}^{N_f} \cos\alpha_i+\frac{m_\pi^4}{4 s\mu^2} \left(\sum_{i=1}^{2s} \cos\alpha_i\right)^2+ s \mu^2 \right) \,, \quad \text{Superfluid phase}\,.
\end{align}
In the Normal phase, the EOM for the Witten variables has the following general solution
\begin{align} \label{Solgen2}
    \alpha_{i}=\begin{cases}
  \pi-  \alpha ,\qquad  & i=1,\dots,n \,, \\
  \alpha ,\qquad & i=n+1,\dots,N_f \,,
    \end{cases} 
\end{align}
where
\begin{equation} \label{Solgen}
    \alpha=\frac{\theta+ (2k-n)\pi}{(N_{f}-2n)},\quad k=0,\dots, N_f-2n-1 \,, \quad n=
    0,..., \left[\frac{N_f-1}{2}\right] \,.
\end{equation}

As shown in \cite{Bersini:2022jhs}, the solution minimizing the energy reads
\be \label{questaequazione}
\alpha=\begin{cases}
    \frac{\theta}{N_f}  \,, \qquad \theta \in [0, \pi] \,, \\
     \frac{\theta -2\pi}{N_f}\,, \ \quad \theta \in [\pi, 2\pi]  \,,
\end{cases} 
\ee
corresponding to $n=0$ and, respectively, $k=0$ and $k=N_f-1$.
We conclude that in the case of degenerate quark masses, the Normal phase features spontaneous breaking of CP symmetry at $\theta=\pi$ for any value of $N_f$.  To discuss the superfluid phase, we make the following reasonable assumption on the form of the ground state
\begin{equation}
\alpha_i=\begin{cases}
    \alpha_1 \,, \quad i=1, \dots 2s  \,, \\
    \alpha_2 \,, \quad i=2s+1, \dots N_f  \,.
\end{cases}     
\end{equation}
Hence for $N_f \neq 2s$ the equations to solve are
\begin{align} \label{eqcompleta}
2\alpha_1 =\frac{\theta-\left(N_f-2s\right)\alpha_2}{s}+\frac{2\pi k}{s} \,, \quad
  \sin\alpha_2  =   \frac{z}{2}  \sin\left(\frac{\theta+2\pi k-\left(N_f-2s\right)\alpha_2}{s}\right)\,,
\end{align}
where $k=0,\dots, 2s-1$ and $z \equiv \frac{m_\pi^2}{\mu^2}$. When $s=N_f/2$, the EOM are the same as in the Normal phase and the $\theta$-vacuum is again given by Eq. \eqref{questaequazione}. In the particular case $s =N_f/3$, Eq. \eqref{eqcompleta} can be solved exactly with the relevant solutions being
\begin{align}
    \alpha_{1} &= \frac{3 \theta }{2 N_f}-\frac{\alpha_2}{2}  \,, \qquad \quad \ \alpha_2 = \arccos \left(\frac{\gamma  N_f \cos \left(\frac{3 \theta }{N_f}\right)+3}{\sqrt{\gamma ^2 N_f^2+6 \gamma  N_f \cos \left(\frac{3 \theta }{N_f}\right)+9}}\right) \,, \nonumber \\   \alpha_{1} &=\frac{ 3 (\theta - 2 \pi)}{2 N_f}-\frac{\alpha_2}{2} \,, \quad \alpha_2 = -\arccos \left(\frac{\gamma  N_f \cos \left(\frac{6 \pi -3 \theta }{N_f}\right)+3}{\sqrt{\gamma ^2 N_f^2+6 \gamma  N_f \cos \left(\frac{6 \pi -3 \theta }{N_f}\right)+9}}\right) \,,
\end{align}
which minimize the energy for $\theta \in [0,\pi]$ and  $\theta \in [\pi,2\pi]$, respectively. The two solutions cross at $\theta = \pi$ where Dashen's phenomenon occurs.

For generic values of $s$, Eq. \eqref{eqcompleta} does not admit a simple analytic solution. However, progress can be made by assuming $z =m_\pi/\mu \ll 1$. In this limit, the ground state solutions are 
\begin{align}
    \alpha_{1,a}&=  \frac{\theta +2 \pi  k}{2 s} - \frac{ N_f-2 s }{4 s} \sin \left(\frac{\theta +2 \pi  k}{s}\right)z + \frac{ (N_f-2 s)^2}{16 s^2} \sin \left(\frac{2 (\theta +2 \pi  k)}{s}\right)z^2 + \cO \left(z^3 \right) \,, \nonumber\\
    \alpha_{2,a}&=  \frac{1}{2}  \sin \left(\frac{\theta +2 \pi  k}{s}\right)z - \frac{N_f-2 s }{8 s}\sin \left(\frac{2 (\theta +2 \pi  k)}{s}\right) z^2 + \cO \left(z^3 \right) \,,  
    % \nonumber \\
    % \alpha_{1,b}&= \frac{\theta +2 \pi  k-\pi  N_f+2 \pi  s}{2 s} + \frac{ N_f-2 s }{4 s} \sin \left(\frac{\theta +2 \pi  k-\pi  N_f}{s}\right) z  \nonumber\\ &+ \frac{ (N_f-2 s)^2}{16 s^2} \sin \left(\frac{2 (\theta +2 \pi  k-\pi  N_f)}{s}\right)z^2 + \cO \left(z^3 \right) \,, \nonumber\\
    % \alpha_{2,b}&= \pi - \frac{1}{2}  \sin \left(\frac{\theta +2 \pi  k-\pi  N_f}{s}\right)z -\frac{N_f-2 s}{8 s} \sin \left(\frac{2 (\theta +2 \pi  k-\pi  N_f)}{s}\right)z^2 + \cO \left(z^3 \right) \,.
\end{align}
with $k=1, \dots, 2 s-1$. The corresponding energy is
\begin{align}
    E_S&= -\frac{F_\pi^2 \mu^2}{2} \Bigg(s +  (N_f-2 s)z + s  \cos ^2\left(\frac{\theta +2 \pi  k}{2 s}\right)z^2 + \frac{1}{8}(N_f-2 s) \sin ^2\left(\frac{\theta +2 \pi  k}{s}\right)  z^3 + \cO \left(z^4 \right) \Bigg) \,.
\end{align}
To determine the value of $k$, we note that $E_S$ is minimized when $\cos ^2\left(\frac{\theta +2 \pi  k}{2 s}\right)$ is maximized. This yields $k=0$ for $\theta \in [0,\pi]$ and $k=s-1$ for $\theta\in [\pi,2\pi]$. We conclude that $CP$ symmetry is spontaneously broken at $\theta=\pi$ for any $s>1$. On the other hand, for $s=1$ the two solutions are degenerate for every value of $\theta$, and Dashen's phenomenon is washed out in the superfluid phase.

\section{Conclusions} \label{conclusions}
\label{Conclusions}

We studied the effects of the topological sector on the low energy QCD phase diagram at nonzero isospin and strangeness chemical potentials. To this end, we employed the effective chiral Lagrangian mostly focusing on the three light flavors case while varying the quark mass ratios. The investigation unveiled the nature of the various phase transitions as well as the dependence of the phase boundaries on the QCD $\theta$-angle. 

In particular, we observed that Dashen's phenomenon is absent in the superfluid phases, and, for $\theta \sim \pi$, we found a range of values of the quark mass ratio for which the superfluid phase transitions occur for tiny values of the chemical potentials. Another noteworthy result is the discovery of a novel parity-preserving superfluid phase that can only be realized at $\theta = \pi$. Finally, to further characterize the phase diagram, we calculated the value of the various condensates and charge densities and determined the $\theta$-dependence of the meson mass spectrum in all the phases. Our results can be of interest for novel models of strongly coupled dynamics and their phase transition in the early universe, including possible imprints on gravitational waves (see, for example \cite{Pasechnik:2023hwv,Reichert:2021cvs,Huang:2020crf,Bruno:2024dha,Shao:2024dxt,Guo:2024gmu,McKeen:2024trt,Wang:2024wcs,Ritter:2024sqv,Roshan:2024qnv,Springer:2023hcc,Fujikura:2023fbi,Freese:2023fcr,Khoze:2022nyt,Mason:2022aka,Morgante:2022zvc,Ares:2021nap}), once extending the phase diagram at finite temperature. It would be interesting to use our results as a testbed for lattice simulation while extending them beyond the domain of validity of the effective chiral Lagrangian.

\section*{Acknowledgments}

The work of J.B. was supported by the World Premier International Research Center Initiative (WPI Initiative), MEXT, Japan, and by the JSPS KAKENHI Grant Number JP23K19047. The work of F.S. is partially supported by the Carlsberg Foundation, grant CF22-0922.


\newpage

\begin{thebibliography}{1}

%\cite{Callan:1976je}
\bibitem{Callan:1976je}
C.~G.~Callan, Jr., R.~F.~Dashen and D.~J.~Gross,
``The Structure of the Gauge Theory Vacuum,''
Phys. Lett. B \textbf{63} (1976), 334-340
doi:10.1016/0370-2693(76)90277-X
%1612 citations counted in INSPIRE as of 22 Dec 2024

%\cite{Rajagopal:2000wf}
\bibitem{Rajagopal:2000wf}
K.~Rajagopal and F.~Wilczek,
``The Condensed matter physics of QCD,''
doi:10.1142/9789812810458\_0043
[arXiv:hep-ph/0011333 [hep-ph]].
%952 citations counted in INSPIRE as of 28 Dec 2024

%\cite{Sannino:2009za}
\bibitem{Sannino:2009za}
F.~Sannino,
``Conformal Dynamics for TeV Physics and Cosmology,''
Acta Phys. Polon. B \textbf{40} (2009), 3533-3743
[arXiv:0911.0931 [hep-ph]].
%313 citations counted in INSPIRE as of 28 Dec 2024

%\cite{Boer:2008ct}
\bibitem{Boer:2008ct}
D.~Boer and J.~K.~Boomsma,
``Spontaneous CP-violation in the strong interaction at theta = pi,''
Phys. Rev. D \textbf{78} (2008), 054027
doi:10.1103/PhysRevD.78.054027
[arXiv:0806.1669 [hep-ph]].

%\cite{Sakai:2011gs}
\bibitem{Sakai:2011gs}
Y.~Sakai, H.~Kouno, T.~Sasaki and M.~Yahiro,
``Theta vacuum effects on QCD phase diagram,''
Phys. Lett. B \textbf{705} (2011), 349-355
doi:10.1016/j.physletb.2011.10.032
[arXiv:1105.0413 [hep-ph]].
%24 citations counted in INSPIRE as of 18 Oct 2024

%\cite{DElia:2013uaf}
\bibitem{DElia:2013uaf}
M.~D'Elia and F.~Negro,
``Phase diagram of Yang-Mills theories in the presence of a $\theta$ term,''
Phys. Rev. D \textbf{88} (2013) no.3, 034503
doi:10.1103/PhysRevD.88.034503
[arXiv:1306.2919 [hep-lat]].
%53 citations counted in INSPIRE as of 18 Oct 2024

%\cite{Chatterjee:2011yz}
\bibitem{Chatterjee:2011yz}
B.~Chatterjee, H.~Mishra and A.~Mishra,
``Strong CP violation and chiral symmetry breaking in hot and dense quark matter,''
Phys. Rev. D \textbf{85} (2012), 114008
doi:10.1103/PhysRevD.85.114008
[arXiv:1111.4061 [hep-ph]].
%8 citations counted in INSPIRE as of 18 Oct 2024

%\cite{Chatterjee:2014csa}
\bibitem{Chatterjee:2014csa}
B.~Chatterjee, H.~Mishra and A.~Mishra,
``$CP$ violation and chiral symmetry breaking in hot and dense quark matter in the presence of a magnetic field,''
Phys. Rev. D \textbf{91} (2015) no.3, 034031
doi:10.1103/PhysRevD.91.034031
[arXiv:1409.3454 [hep-ph]].
%15 citations counted in INSPIRE as of 18 Oct 2024

%\cite{Murgana:2024djt}
\bibitem{Murgana:2024djt}
F.~Murgana, D.~E.~A.~Castillo, A.~G.~Grunfeld and M.~Ruggieri,
``Topological susceptibility and axion potential in two-flavor superconductive quark matter,''
Phys. Rev. D \textbf{110} (2024) no.1, 014042
doi:10.1103/PhysRevD.110.014042
[arXiv:2404.14160 [hep-ph]].
%1 citations counted in INSPIRE as of 24 Nov 2024

%\cite{Son:2000xc}
\bibitem{Son:2000xc}
D.~T.~Son and M.~A.~Stephanov,
``QCD at finite isospin density,''
Phys. Rev. Lett. \textbf{86} (2001), 592-595
doi:10.1103/PhysRevLett.86.592
[arXiv:hep-ph/0005225 [hep-ph]].
%633 citations counted in INSPIRE as of 17 Oct 2024

%\cite{Son:2000by}
\bibitem{Son:2000by}
D.~T.~Son and M.~A.~Stephanov,
``QCD at finite isospin density: From pion to quark - anti-quark condensation,''
Phys. Atom. Nucl. \textbf{64} (2001), 834-842
doi:10.1134/1.1378872
[arXiv:hep-ph/0011365 [hep-ph]].
%242 citations counted in INSPIRE as of 17 Oct 2024

%\cite{Schafer:2001bq}
\bibitem{Schafer:2001bq}
T.~Sch\"afer, D.~T.~Son, M.~A.~Stephanov, D.~Toublan and J.~J.~M.~Verbaarschot,
``Kaon condensation and Goldstone's theorem,''
Phys. Lett. B \textbf{522} (2001), 67-75
doi:10.1016/S0370-2693(01)01265-5
[arXiv:hep-ph/0108210 [hep-ph]].
%166 citations counted in INSPIRE as of 17 Oct 2024

%\cite{Kogut:2001id}
\bibitem{Kogut:2001id}
J.~B.~Kogut and D.~Toublan,
``QCD at small nonzero quark chemical potentials,''
Phys. Rev. D \textbf{64} (2001), 034007
doi:10.1103/PhysRevD.64.034007
[arXiv:hep-ph/0103271 [hep-ph]].
%162 citations counted in INSPIRE as of 17 Oct 2024

%\cite{Kogut:2002zg}
\bibitem{Kogut:2002zg}
J.~B.~Kogut and D.~K.~Sinclair,
``Lattice QCD at finite isospin density at zero and finite temperature,''
Phys. Rev. D \textbf{66} (2002), 034505
doi:10.1103/PhysRevD.66.034505
[arXiv:hep-lat/0202028 [hep-lat]].
%284 citations counted in INSPIRE as of 17 Oct 2024

%\cite{Loewe:2002tw}
\bibitem{Loewe:2002tw}
M.~Loewe and C.~Villavicencio,
``Thermal pions at finite isospin chemical potential,''
Phys. Rev. D \textbf{67} (2003), 074034
doi:10.1103/PhysRevD.67.074034
[arXiv:hep-ph/0212275 [hep-ph]].
%118 citations counted in INSPIRE as of 17 Oct 2024

%\cite{Barducci:2004tt}
\bibitem{Barducci:2004tt}
A.~Barducci, R.~Casalbuoni, G.~Pettini and L.~Ravagli,
``A Calculation of the QCD phase diagram at finite temperature, and baryon and isospin chemical potentials,''
Phys. Rev. D \textbf{69} (2004), 096004
doi:10.1103/PhysRevD.69.096004
[arXiv:hep-ph/0402104 [hep-ph]].
%159 citations counted in INSPIRE as of 17 Oct 2024

%\cite{He:2005nk}
\bibitem{He:2005nk}
L.~y.~He, M.~Jin and P.~f.~Zhuang,
``Pion superfluidity and meson properties at finite isospin density,''
Phys. Rev. D \textbf{71} (2005), 116001
doi:10.1103/PhysRevD.71.116001
[arXiv:hep-ph/0503272 [hep-ph]].
%247 citations counted in INSPIRE as of 17 Oct 2024

%\cite{Andersen:2006ys}
\bibitem{Andersen:2006ys}
J.~O.~Andersen,
``Pion and kaon condensation at finite temperature and density,''
Phys. Rev. D \textbf{75} (2007), 065011
doi:10.1103/PhysRevD.75.065011
[arXiv:hep-ph/0609020 [hep-ph]].
%73 citations counted in INSPIRE as of 17 Oct 2024

%\cite{Adhikari:2019zaj}
\bibitem{Adhikari:2019zaj}
P.~Adhikari and J.~O.~Andersen,
``QCD at finite isospin density: chiral perturbation theory confronts lattice data,''
Phys. Lett. B \textbf{804} (2020), 135352
doi:10.1016/j.physletb.2020.135352
[arXiv:1909.01131 [hep-ph]].
%38 citations counted in INSPIRE as of 17 Oct 2024

%\cite{Adhikari:2019mlf}
\bibitem{Adhikari:2019mlf}
P.~Adhikari and J.~O.~Andersen,
``Pion and kaon condensation at zero temperature in three-flavor $\chi$PPT at nonzero isospin and strange chemical potentials at next-to-leading order,''
JHEP \textbf{06} (2020), 170
doi:10.1007/JHEP06(2020)170
[arXiv:1909.10575 [hep-ph]].
%22 citations counted in INSPIRE as of 20 Dec 2024

%\cite{Detmold:2008yn}
\bibitem{Detmold:2008yn}
W.~Detmold, K.~Orginos, M.~J.~Savage and A.~Walker-Loud,
``Kaon Condensation with Lattice QCD,''
Phys. Rev. D \textbf{78} (2008), 054514
doi:10.1103/PhysRevD.78.054514
[arXiv:0807.1856 [hep-lat]].
%100 citations counted in INSPIRE as of 17 Oct 2024

%\cite{Brandt:2017zck}
\bibitem{Brandt:2017zck}
B.~B.~Brandt, G.~Endrodi and S.~Schmalzbauer,
``QCD at finite isospin chemical potential,''
EPJ Web Conf. \textbf{175} (2018), 07020
doi:10.1051/epjconf/201817507020
[arXiv:1709.10487 [hep-lat]].
%58 citations counted in INSPIRE as of 17 Oct 2024

%\cite{Mammarella:2015pxa}
\bibitem{Mammarella:2015pxa}
A.~Mammarella and M.~Mannarelli,
``Intriguing aspects of meson condensation,''
Phys. Rev. D \textbf{92} (2015) no.8, 085025
doi:10.1103/PhysRevD.92.085025
[arXiv:1507.02934 [hep-ph]].
%35 citations counted in INSPIRE as of 17 Oct 2024

%\cite{Carignano:2016lxe}
\bibitem{Carignano:2016lxe}
S.~Carignano, L.~Lepori, A.~Mammarella, M.~Mannarelli and G.~Pagliaroli,
``Scrutinizing the pion condensed phase,''
Eur. Phys. J. A \textbf{53} (2017) no.2, 35
doi:10.1140/epja/i2017-12221-x
[arXiv:1610.06097 [hep-ph]].
%74 citations counted in INSPIRE as of 17 Oct 2024

%\cite{Mannarelli:2019hgn}
\bibitem{Mannarelli:2019hgn}
M.~Mannarelli,
``Meson condensation,''
Particles \textbf{2} (2019) no.3, 411-443
doi:10.3390/particles2030025
[arXiv:1908.02042 [hep-ph]].
%75 citations counted in INSPIRE as of 20 Dec 2024

%\cite{GomezNicola:2022asf}
\bibitem{GomezNicola:2022asf}
A.~G\'omez Nicola and A.~Vioque-Rodr\'\i{}guez,
``Effective Lagrangian at nonzero isospin chemical potential,''
Phys. Rev. D \textbf{106} (2022) no.11, 114017
doi:10.1103/PhysRevD.106.114017
[arXiv:2205.14609 [hep-ph]].
%5 citations counted in INSPIRE as of 17 Oct 2024

%\cite{Ayala:2023cnt}
\bibitem{Ayala:2023cnt}
A.~Ayala, A.~Bandyopadhyay, R.~L.~S.~Farias, L.~A.~Hern\'andez and J.~L.~Hern\'andez,
``QCD equation of state at finite isospin density from the linear sigma model with quarks: The cold case,''
Phys. Rev. D \textbf{107} (2023) no.7, 074027
doi:10.1103/PhysRevD.107.074027
[arXiv:2301.13633 [hep-ph]].
%12 citations counted in INSPIRE as of 17 Oct 2024

%\cite{Abbott:2024vhj}
\bibitem{Abbott:2024vhj}
R.~Abbott, W.~Detmold, M.~Illa, A.~Parre\~no, R.~J.~Perry, F.~Romero-L\'opez, P.~E.~Shanahan and M.~L.~Wagman,
``QCD constraints on isospin-dense matter and the nuclear equation of state,''
[arXiv:2406.09273 [hep-lat]].
%11 citations counted in INSPIRE as of 17 Oct 2024


%\cite{Metlitski:2005di}
\bibitem{Metlitski:2005di}
M.~A.~Metlitski and A.~R.~Zhitnitsky,
``theta-dependence of QCD at finite isospin density,''
Phys. Lett. B \textbf{633} (2006), 721-728
doi:10.1016/j.physletb.2006.01.001
[arXiv:hep-ph/0510162 [hep-ph]].
%26 citations counted in INSPIRE as of 18 Oct 2024

%\cite{Metlitski:2005db}
\bibitem{Metlitski:2005db}
M.~A.~Metlitski and A.~R.~Zhitnitsky,
``Theta-parameter in 2 color QCD at finite baryon and isospin density,''
Nucl. Phys. B \textbf{731} (2005), 309-334
doi:10.1016/j.nuclphysb.2005.09.027
[arXiv:hep-ph/0508004 [hep-ph]].
%42 citations counted in INSPIRE as of 22 Aug 2022

%\cite{Bersini:2022jhs}
\bibitem{Bersini:2022jhs}
J.~Bersini, A.~D'Alise, F.~Sannino and M.~Torres,
``The \ensuremath{\theta}-angle and axion physics of two-color QCD at fixed baryon charge,''
JHEP \textbf{11} (2022), 080
doi:10.1007/JHEP11(2022)080
[arXiv:2208.09226 [hep-th]].
%2 citations counted in INSPIRE as of 03 Nov 2023

%\cite{Balkin:2020dsr}
\bibitem{Balkin:2020dsr}
R.~Balkin, J.~Serra, K.~Springmann and A.~Weiler,
``The QCD axion at finite density,''
JHEP \textbf{07} (2020), 221
doi:10.1007/JHEP07(2020)221
[arXiv:2003.04903 [hep-ph]].
%41 citations counted in INSPIRE as of 18 Oct 2024

%\cite{Peccei:1977hh}
\bibitem{Peccei:1977hh}
R.~D.~Peccei and H.~R.~Quinn,
``CP Conservation in the Presence of Instantons,''
Phys. Rev. Lett. \textbf{38} (1977), 1440-1443
doi:10.1103/PhysRevLett.38.1440
%8104 citations counted in INSPIRE as of 18 Oct 2024

%\cite{Peccei:1977ur}
\bibitem{Peccei:1977ur}
R.~D.~Peccei and H.~R.~Quinn,
``Constraints Imposed by CP Conservation in the Presence of Instantons,''
Phys. Rev. D \textbf{16} (1977), 1791-1797
doi:10.1103/PhysRevD.16.1791
%3458 citations counted in INSPIRE as of 01 Sep 2022

%\cite{Cacciapaglia:2020kgq}
\bibitem{Cacciapaglia:2020kgq}
G.~Cacciapaglia, C.~Pica and F.~Sannino,
``Fundamental Composite Dynamics: A Review,''
Phys. Rept. \textbf{877} (2020), 1-70
doi:10.1016/j.physrep.2020.07.002
[arXiv:2002.04914 [hep-ph]].
%70 citations counted in INSPIRE as of 23 Aug 2022

%\cite{Kharzeev:1998kz}
\bibitem{Kharzeev:1998kz}
D.~Kharzeev, R.~D.~Pisarski and M.~H.~G.~Tytgat,
``Possibility of spontaneous parity violation in hot QCD,''
Phys. Rev. Lett. \textbf{81} (1998), 512-515
doi:10.1103/PhysRevLett.81.512
[arXiv:hep-ph/9804221 [hep-ph]].
%410 citations counted in INSPIRE as of 23 Nov 2024

%\cite{Voloshin:2004vk}
\bibitem{Voloshin:2004vk}
S.~A.~Voloshin,
``Parity violation in hot QCD: How to detect it,''
Phys. Rev. C \textbf{70} (2004), 057901
doi:10.1103/PhysRevC.70.057901
[arXiv:hep-ph/0406311 [hep-ph]].
%407 citations counted in INSPIRE as of 23 Nov 2024


%\cite{Ruck:1976zt}
\bibitem{Ruck:1976zt}
V.~Ruck, M.~Gyulassy and W.~Greiner,
``Pion Condensation in Heavy Ion Collisions,''
Z. Phys. A \textbf{277} (1976), 391-394
doi:10.1007/BF01545977
%68 citations counted in INSPIRE as of 23 Nov 2024

%\cite{Greiner:1993jn}
\bibitem{Greiner:1993jn}
C.~Greiner, C.~Gong and B.~Muller,
``Some remarks on pion condensation in relativistic heavy ion collisions,''
Phys. Lett. B \textbf{316} (1993), 226-230
doi:10.1016/0370-2693(93)90317-B
[arXiv:hep-ph/9307336 [hep-ph]].
%72 citations counted in INSPIRE as of 23 Nov 2024

%\cite{Migdal:1990vm}
\bibitem{Migdal:1990vm}
A.~B.~Migdal, E.~E.~Saperstein, M.~A.~Troitsky and D.~N.~Voskresensky,
``Pion degrees of freedom in nuclear matter,''
Phys. Rept. \textbf{192} (1990), 179-437
doi:10.1016/0370-1573(90)90132-L
%291 citations counted in INSPIRE as of 23 Nov 2024

%\cite{Dashen:1970et}
\bibitem{Dashen:1970et}
R.~F.~Dashen,
``Some features of chiral symmetry breaking,''
Phys. Rev. D \textbf{3} (1971), 1879-1889
doi:10.1103/PhysRevD.3.1879
%392 citations counted in INSPIRE as of 17 Oct 2024

%\cite{Witten:1980sp}
\bibitem{Witten:1980sp}
E.~Witten,
``Large N Chiral Dynamics,''
Annals Phys. \textbf{128} (1980), 363
doi:10.1016/0003-4916(80)90325-5
%718 citations counted in INSPIRE as of 18 Oct 2024

%\cite{DiVecchia:1980yfw}
\bibitem{DiVecchia:1980yfw}
P.~Di Vecchia and G.~Veneziano,
``Chiral Dynamics in the Large n Limit,''
Nucl. Phys. B \textbf{171} (1980), 253-272
doi:10.1016/0550-3213(80)90370-3
%805 citations counted in INSPIRE as of 18 Oct 2024

%\cite{Smilga:1998dh}
\bibitem{Smilga:1998dh}
A.~V.~Smilga,
``QCD at theta similar to pi,''
Phys. Rev. D \textbf{59} (1999), 114021
doi:10.1103/PhysRevD.59.114021
[arXiv:hep-ph/9805214 [hep-ph]].
%82 citations counted in INSPIRE as of 17 Oct 2024

%\cite{Creutz:1995wf}
\bibitem{Creutz:1995wf}
M.~Creutz,
``Quark masses and chiral symmetry,''
Phys. Rev. D \textbf{52} (1995), 2951-2959
doi:10.1103/PhysRevD.52.2951
[arXiv:hep-th/9505112 [hep-th]].
%94 citations counted in INSPIRE as of 17 Oct 2024

%\cite{Creutz:2003xu}
\bibitem{Creutz:2003xu}
M.~Creutz,
``Spontaneous violation of CP symmetry in the strong interactions,''
Phys. Rev. Lett. \textbf{92} (2004), 201601
doi:10.1103/PhysRevLett.92.201601
[arXiv:hep-lat/0312018 [hep-lat]].
%72 citations counted in INSPIRE as of 18 Oct 2024

%\cite{Gaiotto:2017tne}
\bibitem{Gaiotto:2017tne}
D.~Gaiotto, Z.~Komargodski and N.~Seiberg,
``Time-reversal breaking in QCD$_{4}$, walls, and dualities in 2 + 1 dimensions,''
JHEP \textbf{01} (2018), 110
doi:10.1007/JHEP01(2018)110
[arXiv:1708.06806 [hep-th]].
%217 citations counted in INSPIRE as of 17 Oct 2024

%\cite{DiVecchia:2013swa}
\bibitem{DiVecchia:2013swa}
P.~Di Vecchia and F.~Sannino,
``The Physics of the $\theta$-angle for Composite Extensions of the Standard Model,''
Eur. Phys. J. Plus \textbf{129} (2014), 262
doi:10.1140/epjp/i2014-14262-4
[arXiv:1310.0954 [hep-ph]].
%26 citations counted in INSPIRE as of 22 Dec 2024

%\cite{DiVecchia:2017xpu}
\bibitem{DiVecchia:2017xpu}
P.~Di Vecchia, G.~Rossi, G.~Veneziano and S.~Yankielowicz,
``Spontaneous $CP$ breaking in QCD and the axion potential: an effective Lagrangian approach,''
JHEP \textbf{12} (2017), 104
doi:10.1007/JHEP12(2017)104
[arXiv:1709.00731 [hep-th]].
%43 citations counted in INSPIRE as of 17 Oct 2024


%\cite{Pasechnik:2023hwv}
\bibitem{Pasechnik:2023hwv}
R.~Pasechnik, M.~Reichert, F.~Sannino and Z.~W.~Wang,
``Gravitational waves from composite dark sectors,''
JHEP \textbf{02}, 159 (2024)
doi:10.1007/JHEP02(2024)159
[arXiv:2309.16755 [hep-ph]].
%26 citations counted in INSPIRE as of 07 Jan 2025


%\cite{Reichert:2021cvs}
\bibitem{Reichert:2021cvs}
M.~Reichert, F.~Sannino, Z.~W.~Wang and C.~Zhang,
``Dark confinement and chiral phase transitions: gravitational waves vs matter representations,''
JHEP \textbf{01}, 003 (2022)
doi:10.1007/JHEP01(2022)003
[arXiv:2109.11552 [hep-ph]].
%50 citations counted in INSPIRE as of 07 Jan 2025

%\cite{Huang:2020crf}
\bibitem{Huang:2020crf}
W.~C.~Huang, M.~Reichert, F.~Sannino and Z.~W.~Wang,
``Testing the dark SU(N) Yang-Mills theory confined landscape: From the lattice to gravitational waves,''
Phys. Rev. D \textbf{104}, no.3, 035005 (2021)
doi:10.1103/PhysRevD.104.035005
[arXiv:2012.11614 [hep-ph]].
%99 citations counted in INSPIRE as of 07 Jan 2025



%\cite{Bruno:2024dha}
\bibitem{Bruno:2024dha}
M.~Bruno, N.~Forzano, M.~Panero and A.~Smecca,
``Thermal evolution of dark matter in the early universe from a symplectic glueball model,''
[arXiv:2410.17122 [hep-ph]].
%3 citations counted in INSPIRE as of 07 Jan 2025


%\cite{Shao:2024dxt}
\bibitem{Shao:2024dxt}
J.~Shao, H.~Mao and M.~Huang,
``The Transition Rate and Gravitational Wave Spectrum from First-Order QCD Phase Transitions,''
[arXiv:2410.06780 [hep-ph]].
%2 citations counted in INSPIRE as of 07 Jan 2025

%\cite{Guo:2024gmu}
\bibitem{Guo:2024gmu}
H.~k.~Guo, F.~Hajkarim, K.~Sinha, G.~White and Y.~Xiao,
``A Precise Fitting Formula for Gravitational Wave Spectra from Phase Transitions,''
[arXiv:2407.02580 [hep-ph]].
%4 citations counted in INSPIRE as of 07 Jan 2025

%\cite{McKeen:2024trt}
\bibitem{McKeen:2024trt}
D.~McKeen, R.~Mizuta, D.~E.~Morrissey and M.~Shamma,
``Dark Matter from Dark Glueball Dominance,''
[arXiv:2406.18635 [hep-ph]].
%5 citations counted in INSPIRE as of 07 Jan 2025


%\cite{Wang:2024wcs}
\bibitem{Wang:2024wcs}
D.~W.~Wang, Q.~S.~Yan and M.~Huang,
``Bubble wall velocity and gravitational wave in the minimal left-right symmetric model,''
Phys. Rev. D \textbf{110}, no.7, 076011 (2024)
doi:10.1103/PhysRevD.110.076011
[arXiv:2405.01949 [gr-qc]].
%1 citations counted in INSPIRE as of 07 Jan 2025

%\cite{Ritter:2024sqv}
\bibitem{Ritter:2024sqv}
A.~C.~Ritter and R.~R.~Volkas,
``Explaining the cosmological dark matter coincidence in asymmetric dark QCD,''
Phys. Rev. D \textbf{110}, no.1, 015032 (2024)
doi:10.1103/PhysRevD.110.015032
[arXiv:2404.05999 [hep-ph]].
%4 citations counted in INSPIRE as of 07 Jan 2025

%\cite{Roshan:2024qnv}
\bibitem{Roshan:2024qnv}
R.~Roshan and G.~White,
``Using gravitational waves to see the first second of the Universe,''
[arXiv:2401.04388 [hep-ph]].
%48 citations counted in INSPIRE as of 07 Jan 2025

%\cite{Springer:2023hcc}
\bibitem{Springer:2023hcc}
F.~Springer \textit{et al.} [Lattice Strong Dynamics (LSD)],
``First-order bulk transitions in large-$N$ lattice Yang--Mills theories using the density of states,''
[arXiv:2311.10243 [hep-lat]].
%6 citations counted in INSPIRE as of 07 Jan 2025



%\cite{Fujikura:2023fbi}
\bibitem{Fujikura:2023fbi}
K.~Fujikura, Y.~Nakai, R.~Sato and Y.~Wang,
``Cosmological phase transitions in composite Higgs models,''
JHEP \textbf{09}, 053 (2023)
doi:10.1007/JHEP09(2023)053
[arXiv:2306.01305 [hep-ph]].
%11 citations counted in INSPIRE as of 07 Jan 2025


%\cite{Freese:2023fcr}
\bibitem{Freese:2023fcr}
K.~Freese and M.~W.~Winkler,
``Dark matter and gravitational waves from a dark big bang,''
Phys. Rev. D \textbf{107}, no.8, 083522 (2023)
doi:10.1103/PhysRevD.107.083522
[arXiv:2302.11579 [astro-ph.CO]].
%28 citations counted in INSPIRE as of 07 Jan 2025


%\cite{Khoze:2022nyt}
\bibitem{Khoze:2022nyt}
V.~V.~Khoze and D.~L.~Milne,
``Gravitational waves and dark matter from classical scale invariance,''
Phys. Rev. D \textbf{107}, no.9, 095012 (2023)
doi:10.1103/PhysRevD.107.095012
[arXiv:2212.04784 [hep-ph]].
%18 citations counted in INSPIRE as of 07 Jan 2025



%\cite{Mason:2022aka}
\bibitem{Mason:2022aka}
D.~Mason, B.~Lucini, M.~Piai, E.~Rinaldi and D.~Vadacchino,
``The density of state method for first-order phase transitions in Yang-Mills theories,''
PoS \textbf{LATTICE2022}, 216 (2023)
doi:10.22323/1.430.0216
[arXiv:2212.01074 [hep-lat]].
%12 citations counted in INSPIRE as of 07 Jan 2025



%\cite{Morgante:2022zvc}
\bibitem{Morgante:2022zvc}
E.~Morgante, N.~Ramberg and P.~Schwaller,
``Gravitational waves from dark SU(3) Yang-Mills theory,''
Phys. Rev. D \textbf{107}, no.3, 036010 (2023)
doi:10.1103/PhysRevD.107.036010
[arXiv:2210.11821 [hep-ph]].
%49 citations counted in INSPIRE as of 07 Jan 2025

%\cite{Ares:2021nap}
\bibitem{Ares:2021nap}
F.~R.~Ares, O.~Henriksson, M.~Hindmarsh, C.~Hoyos and N.~Jokela,
``Gravitational Waves at Strong Coupling from an Effective Action,''
Phys. Rev. Lett. \textbf{128}, no.13, 131101 (2022)
doi:10.1103/PhysRevLett.128.131101
[arXiv:2110.14442 [hep-th]].
%32 citations counted in INSPIRE as of 07 Jan 2025


\end{thebibliography}
\end{document}